\documentclass[aip, cha, reprint
 amsmath,amssymb,
 reprint,%
]{revtex4-1}
\usepackage{graphicx}
\usepackage{subfig}
\usepackage{xcolor}
\usepackage{bm}       
\usepackage{amssymb} 
\usepackage{amsmath}
\usepackage{amsfonts}
\usepackage{hyperref}
\usepackage[normalem]{ulem} 
\usepackage{natbib}
\usepackage{float}

\usepackage[utf8]{inputenc}
\usepackage[T1]{fontenc}
\usepackage{etoolbox}
\usepackage{newtxmath}

\makeatletter
\def\@email#1#2{%
 \endgroup
 \patchcmd{\titleblock@produce}
  {\frontmatter@RRAPformat}
  {\frontmatter@RRAPformat{\produce@RRAP{*#1\href{mailto:#2}{#2}}}\frontmatter@RRAPformat}
  {}{}
}%
\makeatother

\begin{document}

\title{Path integral approach for predicting the diffusive statistics of geometric phases in chaotic Hamiltonian systems}

\author{Ana Silva}
\affiliation{QuTech and Kavli Institute of Nanoscience, Delft University of Technology, 2628 CJ Delft, the Netherlands}
\affiliation{Department of Physics of Complex Systems, Weizmann Institute of Science, Rehovot 76100, Israel}
\author{Efi Efrati}
\email{efi.efrati@weizmann.ac.il}
\affiliation{Department of Physics of Complex Systems, Weizmann Institute of Science, Rehovot 76100, Israel}

\date{\today}

\begin{abstract}
From the integer quantum Hall effect, to swimming at low Reynolds number, geometric phases arise in the description of many different physical systems. In many of these systems the temporal evolution prescribed by the geometric phase can be directly measured by an external observer. By definition, geometric phases rely on the history of the system's internal dynamics, and so their measurement is directly related to temporal correlations in the system. They, thus, provide a sensitive tool for studying chaotic Hamiltonian systems. 
In this work we present a toy model consisting of an autonomous, low-dimensional, chaotic Hamiltonian system designed to have a simple planar internal state space, and a single geometric phase. The diffusive phase dynamics in the highly chaotic regime is thus governed by the loop statistics of planar random walks. We show that the na\"ive loop statistics result in ballistic behavior of the phase, and recover the diffusive behavior by considering a bounded shape space, or a quadratic confining potential.   
\end{abstract}

\maketitle

\begin{quotation}
A non-holonomic constraint relates the dynamics of a constrained variable to the values and evolution of the remaining variables in path dependent fashion. Consequently, the value of the non-holonomically constrained variable depends explicitly on the path history of the remaining degrees of freedom; it is often referred to as a geometric phase. Such path dependence between variables is shared by many seemingly distinct physical systems, from the integer quantum Hall effect, to the cat righting maneuver.

We employ path integral formulation to describe the dynamical behavior of the non-holonomically constrained variable in the highly chaotic regime. We begin by constructing a simplified toy model representing a mixed Hamiltonian system (capable of both regular and chaotic behavior). The model is designed to display a single geometric phase, expressible as planar area elements. 
By appropriately modifying the well known results for random walks, and identifying mathematical analogues in the path integral description of a particle in a magnetic field bounded by a harmonic potential, we are able to recover both the ballistic short time evolution of the geometric phase, and its diffusive long time statistics. 
\end{quotation}

\section{Introduction}
When in a system the dynamics of a dependent variable is expressed in terms of the independent degrees of freedom in a non-integrable way, we say that the dependent variable is non-holonomically constrained~\cite{mc_cauley,goldstein}. In such cases, following a cyclic deformation, the return of the independent variables to their initial states does not imply the restoration of the dependent variables to their original values.
This difference is captured and quantified by a geometric phase~\cite{shap_wilc,berry_phase1,wsgeometric,WilcZee}.
The term geometric arises because the acquired phase depends solely on the properties of the path traced by the cyclic evolution, and not, for instance, on how fast the path is transversed~\cite{wsgeometric}.

 Examples of such phenomena are abundant in physics, and are often  treated in the context of gauge theories~\cite{montgomery,molgeophase}, with perhaps the most well known one being the Berry phase~\cite{berry_phase1,berry_phase2}. 
Classical analogues include the motion of a swimmer through a fluid at low Reynolds number \cite{shap_wilc}, the cyclic change in the state of polarisation of a beam of light~\cite{light_pol}, and the cat righting maneuver that allows cats dropped from rest to reorient their bodies in mid-air to land on their feet \cite{montgomery,square_cat}. 

The last example illustrates the capacity of isolated deformable bodies to undergo zero angular momentum rotations.
The key to allowing such rotations, despite the vanishing of angular momentum, lays in the ambiguity in determining the relative orientation of two dissimilar realizations of a deformable body~\cite{wsgeometric}. Unless the two shapes are congruent, the relative orientation of the two depends on a gauge choice prescribing, for each, a specific "upright" orientation. This allows to explain the cat righting maneuver through a gauge field approach~\cite{montgomery}. Mathematically, this ambiguity in determining orientations arises through the non-holonomicity of the  angular momentum conservation equation~\cite{montgomery,montgomery2}. The toy model we present and study here also belongs to the class of isolated deformable systems and thus can reorient itself with no angular momentum.

A holonomic constraint on the dynamical variables of a system relates its velocities and position coordinates in an integrable fashion, i.e. allowing to integrate the constraint into a relation containing only the coordinates, and independent of the velocities~\cite{goldstein}. A non-holonomic constraint is non-integrable, i.e. cannot be integrated to yield a relation that solely depends on the position coordinates. The angular momentum conservation condition in deformable systems is non-holonomic. 
Therefore, for a deformable system  with vanishing angular momentum, while the angular velocity is completely determined by the dynamics of the independent variables, integrating the angular velocity to obtain the angle itself is path dependent; 
i.e, the angle depends on the history of the system. While the integral prescribing the angle is not independent of the path in phase space, it is independent of the rate at which this path is traversed, depending only on the precise sequence of shape deformations, and thus can be interpreted as a geometric phase~\cite{WilcZee,wsgeometric}.

The isolated harmonic three body system is comprised of three identical masses, connected by three identical springs with a non vanishing rest length, and laying on a frictionless plane. 
A recent study of this system discovered its underlying chaotic dynamics for moderate energy excitations~\cite{ori1,ori2}. The statistics of the non-holonomically constrained orientation of the three-mass system showed three distinct regimes: (i) a low energy regime in which most trajectories are regular and quasi-periodic, (ii) a moderate energy regime in which the system is highly chaotic and the orientation variable displays a diffusive like behavior $\langle\Delta \theta ^2\rangle\propto t$, and (iii) an intermediate energy regime exhibiting a mixed phase space and fractional angular random walk,  $\langle\Delta \theta ^2\rangle\propto t^\alpha$, where $1<\alpha<2$ connecting the two~\cite{ori1,ori2}.

The Hamiltonian of the harmonic three mass system could be dimensionally reduced to depend only on the three shape variables determining the side-lengths of the mass triangle and their respective momenta. When these coordinates are appropriately chosen~\cite{iwai, montgomery,montgomery2}, the evolution of the orientation can be interpreted as summing the solid angles resulting from the projection of trajectories from a three dimensional shape space onto a normalized section of it called the shape sphere. The theoretical study of the statistics of the angular variable is made difficult both by the projection from three dimensions to the surface of the shape-sphere, and by the non-trivial geometric structure of this surface.

In this work we propose and study a variant of this system whose shape space has reduced dimensionality, requires no further projection, and is naturally planar. Consequently, the corresponding geometric phase can be reduced to an area integral of planar random walks, which have been extensively studied~\cite{levy,edwards_stat,brereton1987topological,khandekar1988distribution,duplantier1989areas}, and be subject to further analysis. An additional advantage of the lower dimensionality is the ability to efficiently construct Poincar\'{e} sections. These allow to further study the structure of phase space as the energy is increased and visualise the gradual destruction of regular regions, along with the growth of chaotic regions overtaking increasing portions of phase space.

For isolated chaotic Hamiltonian systems we expect the phase space variables to always be bounded, constrained by the total energy of the system from obtaining arbitrarily large values. Consequently, the statistics displayed by non-holonimically constrained variables in the chaotic regime are expected to follow the diffusive behaviour described here and explained through the appropriately bounded path integrals~\cite{desbois1992algebraic}. 

\section{A toy model}

\begin{figure}[tb]
 \centering
 \includegraphics[width=0.68\columnwidth]{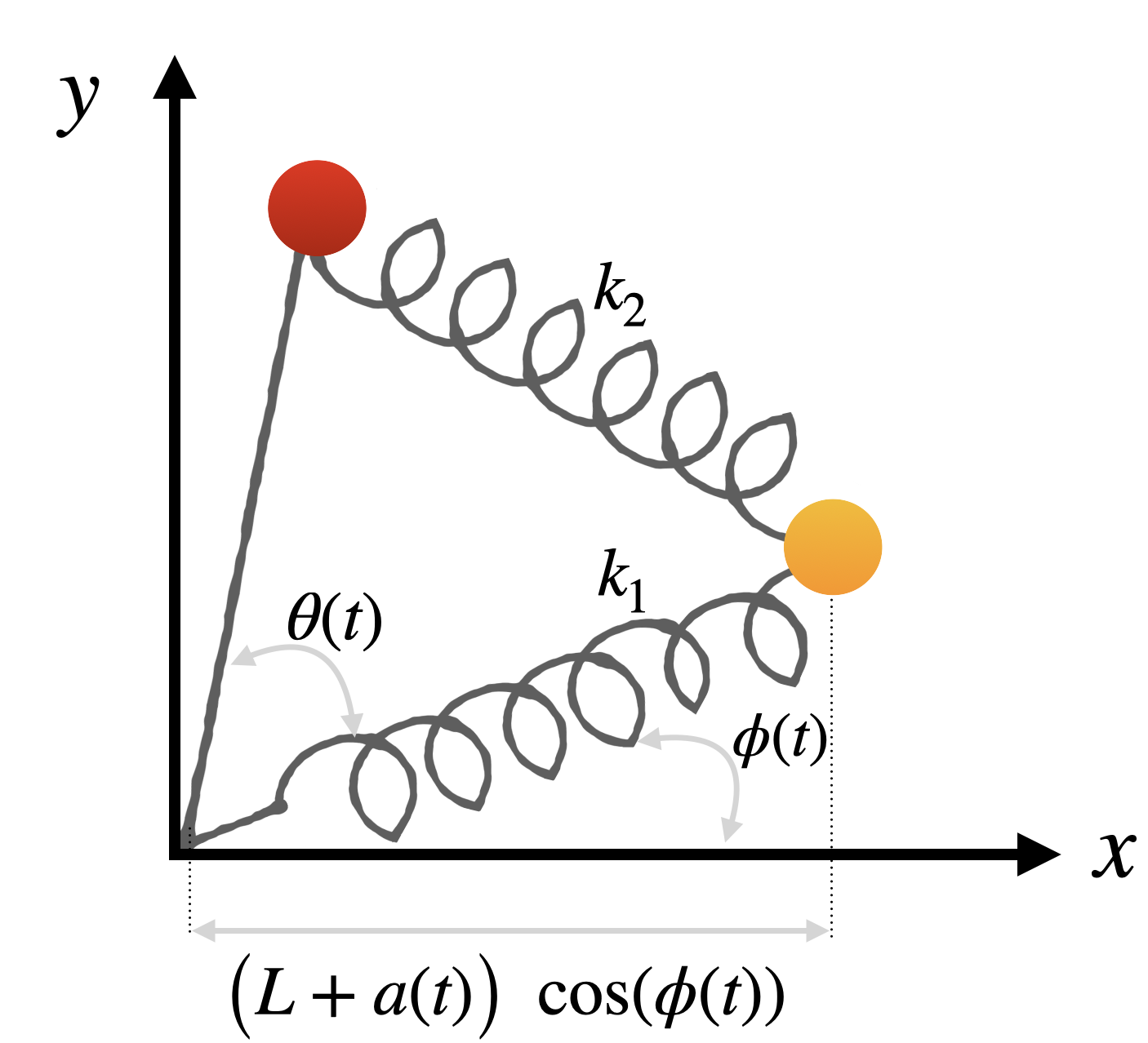}
\caption{The system is made of two equal masses, two springs and a massless rod. The springs have the same rest length $L$, with spring constants, $k_1$ and $k_2$, respectively. The rod has a fixed length $b$, and connects, along with the spring with constant $k_1$, to a hinge. The system has a triangular shape, characterised by the dynamical angle $\theta$, and the length $a$. The configuration of the system is defined with respect to the inertial frame with origin at the hinge, and the orientation is measured by the angle $\phi$ in the figure. \label{SimpleDIR}}
\end{figure}

The toy model consists of a triangular shaped system comprised of two identical masses $m$ connected to each other and to and a frictionless hinge with two harmonic springs and one rigid rod, as depicted in fig.(\ref{SimpleDIR}). The hinge is fixed at the origin of an inertial frame, and joins one of the masses with a rigid rod and the other mass with a harmonic spring.  
A second spring connects the two masses to each other. The entire system is constrained to move in the $2D$ plane. The rod is free to rotate about its hinges but conserves it length, while the springs are also free to rotate, yet can also change their length. \par
We take the rod's length to be fixed and equal to $b$. The springs are assumed to have the identical rest lengths $L$, and spring constants $k_1=k_2=k$ as indicated in the fig.(\ref{SimpleDIR}). This defines $m,k_1,k_2,L,b$ to be the set of constant parameters describing the system. The positions of the two masses in the plane yield four degrees of freedom. The fixed length of the rod reduces these to only three degrees of freedom. At each instance of time, the exact triangular shape can be specified by knowing the internal angle between the rod and the spring $k_1$, $\theta(t)$, and the dynamical length of that spring, which we denote by $a(t)$ (see fig.(\ref{SimpleDIR})). Thus, the set $(a,\theta)$ constitutes the set of shape variables, uniquely defining the shape of the triangle. To fully describe the system we supplement these variables with one orientation variable $\phi$ that determines the orientation of the system in space by prescribing the angle the $k_1$ spring forms with the $x-$axis, (see fig.(\ref{SimpleDIR})). \par 

At each instant of time, the dynamical state of the system is then fully specified by the shape and orientational variables, i.e. the set $(a,\theta,\phi)$. We note that choosing to measure the orientation with respect to the inertial frame's $x-$axis is a gauge choice. This makes the value of the orientational variable $\phi(t)$ gauge dependent. Nevertheless, any changes in orientation measured between equally shaped configurations of the system yield the same difference in orientation, regardless of our particular gauge choice. This ambiguity in prescribing the orientation naturally leads to a gauge structure over the space of shapes~\cite{wsgeometric,montgomery,molgeophase}. In what follows, we adopt the equivalent, but somewhat simpler formulation of the rotation as a geometric phase.  
\\
\subsection{The geometric phase}
Writing the Lagrangian with respect to the variables $a,\theta$ and $\phi$ yields:
\begin{equation}
\mathcal{L}= \frac{m}{2} \Big( \dot{a}^2 + (a+L)^2 \dot{\phi}^2 + b^2 (\dot{\theta}+\dot{\phi})^2 \Big) + V_{12}(a) + V_{23}(a,\theta) \; .
\label{eq:lagrangian}
\end{equation}
The spring potentials $V_{12}(a)$ and $V_{23}(a,\theta)$ are given by

\begin{equation}
\begin{split}
&V_{23}(a,\theta)=\frac{k_2}{2} \Big( \sqrt{b^2+(L+a)^2 -2b(L+a)\cos(\theta)} - L \Big)^2, \\
&V_{12}(a) = \frac{k_1}{2} \Big(|a +L|-L\Big)^2 \end{split}
\label{spring_potentials}
\end{equation}
The system conserves angular momentum ($L_z$), as can be easily proved by observing that $\phi$ is a cyclic coordinate in the Lagrangian. Expressing the conserved angular momentum in terms of $\dot{\phi}$ and $\dot{\theta}$ results in an equation that restricts the co-variation of these variables. In what follows, we will further restrict our attention to the class of systems of vanishing angular momentum, which yield the following relation between rotations and shape deformations:
\begin{equation}
\begin{split}
&L_z=0 \Leftrightarrow \dot{\phi}=-\; \frac{b^2}{b^2+(L+a)^2} \dot{\theta} \\ &\implies \Delta \phi = - \int \frac{b^2}{b^2+(a+L)^2} \; d\theta \; .
\end{split}
\label{def_delta_phi}
\end{equation}
It is clear from eq.(\ref{def_delta_phi}) that $\Delta \phi$ depends on the path followed in shape space yet it is independent of the time parameterisation, thus rendering it a geometric phase~\cite{wsgeometric}. When $\Delta \phi$ represents the change in orientation between two configurations that have the same shape, the path integrated in eq.(\ref{def_delta_phi}) forms a closed loop in shape space. In that case, Stokes' theorem allows us to write:
\begin{equation}
\Delta \phi = \iint \frac{2(a+L)b^2}{\big((a+L)^2+b^2\big)^2} \; dad\theta.
\label{def_delta_phi_2}
\end{equation}
We note that under the transformation
\[
\tilde{a}=-b^2/\big(b^2+(a+L)^2\big),\quad \tilde{\theta}=\theta,
\]
the equation for the change in orientation takes the form
\[
\Delta \phi =\iint d\tilde{a}d\tilde{\theta},
\]
i.e.
the phase $\Delta \phi$ accumulated by a cyclic deformation is given with the area enclosed by the corresponding path in the  $(\tilde{a},\tilde{\theta})$ plane. 
This interpretation will become important later on, when studying the angular mean-square displacement  $\langle \big(\phi(T)-\phi(0) \big)^2\rangle$ , by means of the loop area statistics.

\subsection{The reduced Hamiltonian and its dynamics}
The Hamiltonian of the system in fig.(\ref{SimpleDIR}) associated with the Lagrangian \eqref{eq:lagrangian} is given by
\begin{equation}
\begin{split}
\mathcal{H}= &\frac{1}{2m} \Bigg( p^2_a + \frac{p^2_{\theta}}{b^2} + \frac{(p_{\theta}-p_{\phi})^2}{(L+a)^2}  + \frac{p_{\theta}p_{\phi}}{(L+a)^2} \Bigg) \\ & + V_{12}(a) + V_{23}(a,\theta) \; ,
\end{split}
\end{equation}
with $V_{12}(a)$ and $V_{23}(a,\theta)$ given as in eq.(\ref{spring_potentials}). It is clear that $p_{\phi}$ is a conserved quantity, which is nothing more than the conservation of angular momentum $L_z$. Using eq.(\ref{def_delta_phi}), the Hamiltonian can be re-written solely in terms of the shape variables $(a,\theta)$ (and corresponding conjugate momenta). For the case of vanishing angular momentum the Hamiltonian reads: 
\begin{equation}
\mathcal{H}_{red}= \frac{1}{2m} \Bigg( p^2_a + \frac{p^2_{\theta}}{f(a)} \Bigg) + V_{12}(a) + V_{23}(a,\theta) \; ,
\label{Hred}
\end{equation}
with $f(a)\equiv\frac{b^2(L+a)^2}{b^2+(L+a)^2}$. We refer to this  as the reduced Hamiltonian \cite{symplectic-reduction}.
The reduced Hamiltonian is autonomous and thus conserves energy, reducing its four dimensional dynamical phase space $(a,\theta,p_a,p_{\theta})$, to three dimensions. While the springs' potentials are harmonic, the geometric non-linearities in the system suffice to generate chaos, assuring no other integrals of motion exist~\cite{ott} \cite{strogratz} (see appendix~\ref{app:lle}). We note that chaos in three dynamical dimensions is qualitatively different compared with chaos in higher dimensions, as will be evident soon when visualizing the Poincar\'e sections. 
\begin{figure*}
\includegraphics[width=0.9\textwidth]{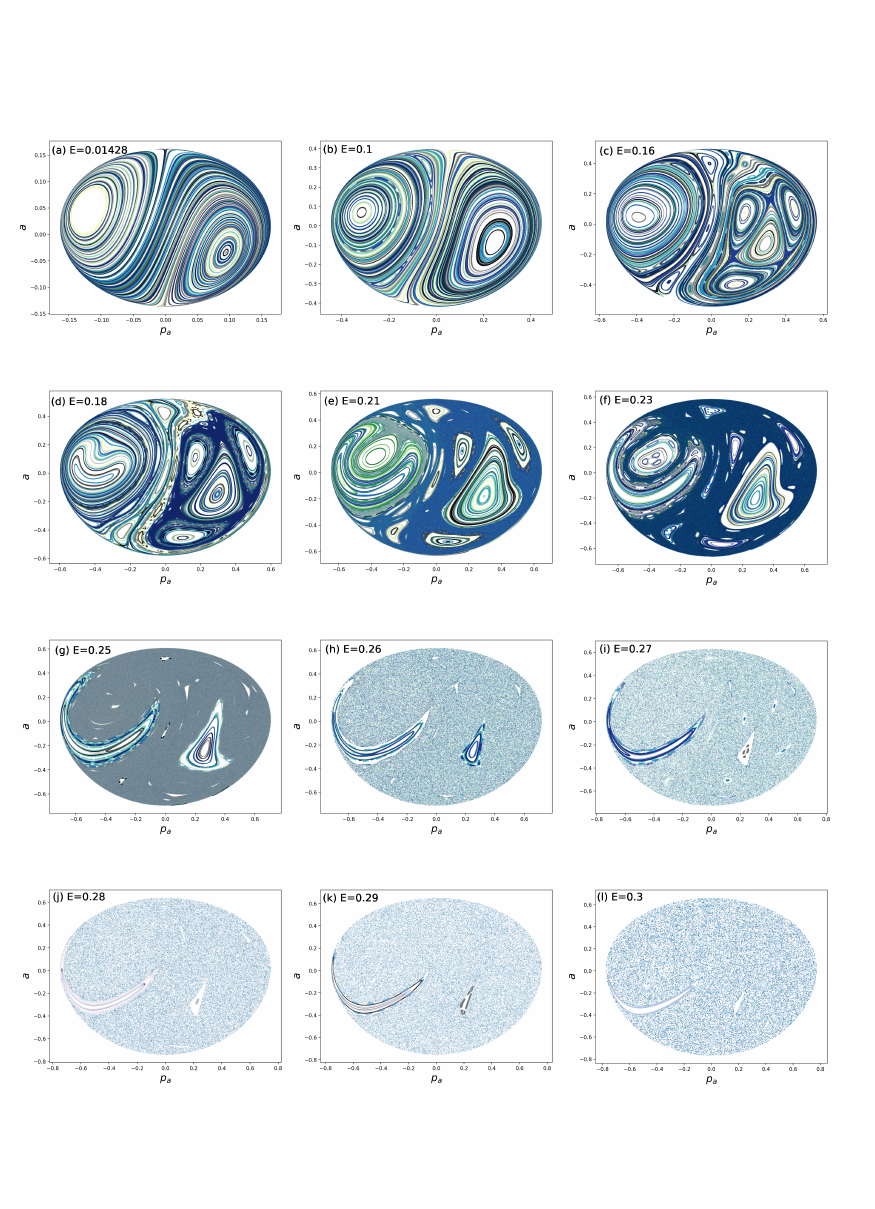}
  \caption{Poincar\'{e} sections for various energies $E$, with the system's parameters chosen to be $m=1$, $k_1=k_2=k=1$ and $L=b=1$. (a): $E=0.0141328$, (b): $E=0.1$, (c): $E=0.16$, (d): $E=0.18$, (e): $E=0.21$, (f): $E=0.23$, (g): $E=0.25$, (h): $E=0.26$, (i): $E=0.27$, (j): $E=0.28$, (k): $E=0.29$ and (l): $E=0.3$.}
   \label{PS}
\end{figure*}

Choosing the system's parameters to read $m=1$, $k_1=k_2=k=1$ and $L=b=1$, a natural reference energy scale emerges $E_{c}=\frac{1}{2}kL^2=\frac{1}{2}$. This is the energy required to simultaneously superimpose the two masses to a single point, and collapse the triangle into a straight line (or equivalently, the energy required to induce a $100\%$ strain in the spring $k_2$). Collinear configurations can be reached whenever $\theta=0$, or $\theta=\pm \pi$. With the previous choice of parameters, the minimal energy $E$ required to reach collinearity is $E_{\text{col}}=\frac{kL^2}{4}=\frac{1}{4}$. Considering energy values smaller than $E_{\text{col}}$, the three mass triangle cannot 
undergo orientation reversals. For energy values much smaller than $E_{\text{col}}$ the system's Hamiltonian may be expanded about the equilibrium equilateral configuration $(a_{\text{min}}=0,\theta_{\text{min}}=\pi/3)$, and the obtained quasiperiodic motion is associated with two non-commensurate frequencies. We note that the system's parameters may be fine-tuned to yield resonance, yet this is not the generic case.

Fig.(\ref{PS}) shows a few Poincar\'{e} sections corresponding to different energies, and assuming our particular choice of parameters. The sections are defined as the plane $\theta=1$, supplemented by a positive value of $p_{\theta}$. This makes $(a,p_a)$ the coordinate system for any of the sections. Each figure is built from a set of initial conditions belonging to the desired section, i.e. they are represented by the vector $(a^{(0)},p^{(0)}_a, 1,p^{(0)}_{\theta})$, with $p^{(0)}_{\theta}>0$. In particular, the values of $(a^{(0)},p^{(0)}_a)$ were chosen at random, while $p^{(0)}_{\theta}$ was determined by the condition that the system has the chosen energy value $E$. Given then the set of initial conditions, the equations of motion were solved numerically using an explicit, symplectic time-reversible second-order integrator~\cite{symplectic_int}, and were evaluated for a long time (up to $\sim10^{6}$ typical oscillation times). We note that while the Hamiltonian we treat is non-separable, the hierarchical structure of its kinetic term allows us to use an integrator developed to solve many-body systems in uniformly curved spaces~\cite{symplectic_int}. 

Whenever the trajectories returned to the plane $\theta=1$ with $p_{\theta}>0$, the corresponding values $(a,p_a)$ were recorded. Thus, each point in fig.(\ref{PS})  represents one of such intersections. All figures were generated from a set of $200$ initial conditions. 

At low energies almost all of phase space is foliated by tori representing quasi periodic trajectories, with no exceptions visible. As each trajectory is restricted to a torus, its crossings of the Poincar\'e section form filamentous structures. As the energy increases, as predicted by KAM theory, the most unstable tori are destroyed and trajectories that occupied these tori and their immediate vicinity are no longer restricted to evolve quasi periodically along the surface of the tori. Consequently, the crossings of the Poincar\'e section of a single trajectory occupy a finite area, rather than concentrate along a line. These chaotic regimes are clearly visible already in fig.(\ref{PS}), $(c)$. 
Note, that the effective dynamical phase space of the system is three dimensional, and as is well known, in three dimensions tori separate space; i.e. two chaotic domains bounded by invariant tori remain disconnected until all separating tori are destroyed. This is clearly visible in fig.(\ref{PS}), $(d)$, and $(e)$ where while some chaotic domains grow and merge, others, separated by invariant tori remain disjoint.

\subsection{Calculating the MSD
in the chaotic regime using loop statistics}
We now examine the angular mean squared displacement (MSD), $\langle\big(\phi(T)-\phi(0)\big)^2\rangle$, in the region where the chaotic behavior prevails. We seek to characterize the statistical behavior of the MSD, and provide an analytical approach that recovers the observed statistics. 

\subsubsection{Numerical analysis of the MSD}
The non-holonomically constrained orientational variable $\phi$, as defined above, can be directly measured as the system evolves in time. As the evolution of $\phi$ depends on the cumulative sequence of shape deformations undergone by the system, it provides a sensitive tool for assessing temporal correlations in the system's internal dynamics and is capable of manifesting exotic power laws~\cite{ori1, ori2}. Note, however, that different choices can be made to measure the orientation of the system. Moreover, the equations that relate the temporal evolution of the orientation of the system to the dynamics of its internal independent degrees of freedom would be different for different choices of the orientation variable. This gauge freedom is not unique to the system at hand and plagues all non-holonomic systems \cite{ori1,wsgeometric}. While in general two distinct states of the system would be associated with different amounts of rotations by the different gauge choices, in some cases all gauge choices must agree. Examining the rotation of the system resulting from a cyclic deformation that starts and ends at the same state of the internal independent variables, leads to such a gauge independent amount of rotation. The existence of such gauge independent measurements allows deducing general conclusions by examining the behavior of particular gauge choices. In what follows we will discuss solely the temporal evolution of $\phi$ as described above.\\  
\\

\begin{figure}[h]
 \centering
 \includegraphics[width=1.1\columnwidth]{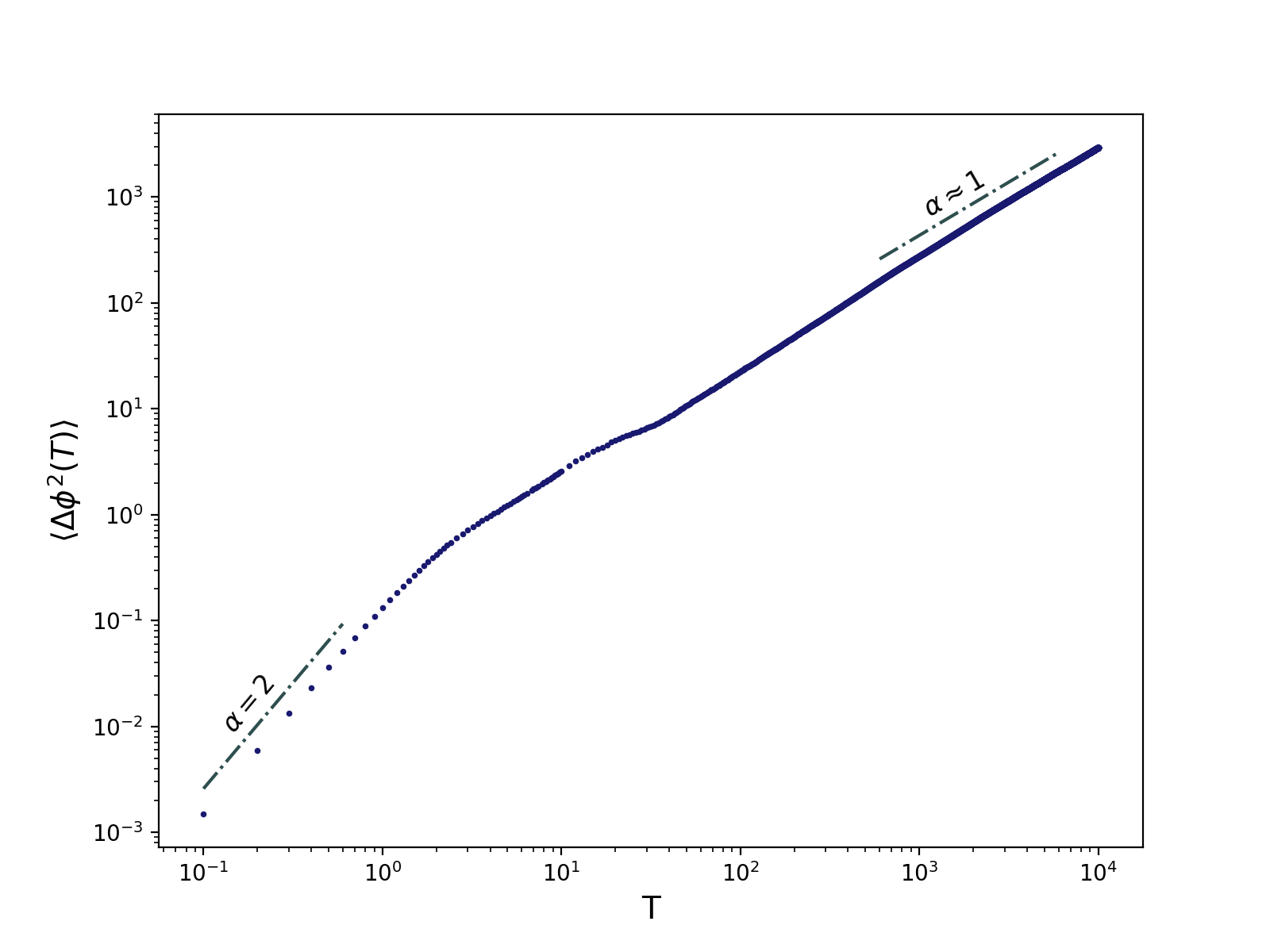}
\caption{Angular mean square displacement (MSD) versus time, $T$, represented in a log-log plot. The displayed MSD was constructed according to eq.(\ref{AMSD_eq}), using an ensemble of 100 initial conditions (IC), all with the same energy $E=0.3$. The green dotted lines serve as a guide to eye, indicating the best fit exponent for the power law $\langle \Delta \phi^2 (T) \rangle \propto T^{\alpha}$. The linear trend occurs when $T$ is well above the typical oscillation time ($T\sim 1$), while for $T<1$, we obtain $\langle \Delta \phi^2 (T) \rangle \propto T^2$. \label{AngularMSD1}}
\end{figure}

In fig.(\ref{AngularMSD1}), we show the numerical evaluation of the angular MSD, $\langle\Delta \phi^2(T)\rangle$ for the fixed set of parameters $\{m,k,b,L\}$ as noted above, at an energy value of $E=0.3$. The angular MSD is calculated through the following average procedure~\cite{ori1}:
\begin{equation}
    \langle\Delta \phi^2(T)\rangle=\frac{1}{p v} \sum^{p}_{i=1} \sum^{v}_{m=1} \Big( \phi_{i,m}(T)-\phi_{i,m}(0) \Big)^2 \; .
    \label{AMSD_eq}
\end{equation}
The sum over $p$ runs over different initial conditions, each sliced temporally to $v$ different time domains. More precisely,
\[
\phi_{i,m}(T)=\phi_{i}(T+(m-1)\tau),
\]
where $\phi_{i}$ is the trajectory that starts from the $i^{th}$ initial condition and is integrated until $t_{max}=10^7$. A total of 122 initial conditions were randomly generated. This procedure relied on sampling the values of $a^{(0)}$, $p^{(0)}_a$, $\theta^{(0)}$ and $\phi^{(0)}$ at random, from a  uniform distribution, with $a^{(0)}$ and $p^{(0)}_a$ sampled over the interval $[-0.8,0.8[$, while $\theta^{(0)}$ and $\phi^{(0)}$ were drawn from the interval $[-\pi,\pi[$. The choice for $p^{(0)}_{\theta}$ was fixed accordingly to guarantee that the generated set of phase space coordinates corresponded to the desired energy ($E=0.3$). Out of the 122 initial conditions, 4 gave rise to regular trajectories, and were excluded from the set. From the remaining 118 trajectories, the first 100 were selected to define our energy ensemble. Each trajectory was then split temporally to $v=\frac{t_{\text{max}}}{\tau}$ different slices, within each $T \in [0,\tau]$. The choice for $\tau$ was set to $\tau=10^4$. Viewing each temporal slice as a different trajectory, the total number of trajectories collected in the double sum, eq.(\ref{AMSD_eq}), is then $pv=10^2\cdot 10^3=10^5$. Figure (\ref{AngularMSD1}) shows the angular MSD for the energy ensemble $E=0.3$. The numerical integration is carried out using  a time-reversible, second order, symplectic solver with a constant time step of $\Delta t=0.01$~\cite{symplectic_int}.

Initially, the numerically evaluated angular MSD shows ballistic propagation, i.e. $\langle\Delta \phi^2(T)\rangle\; \propto \; T^2$~(see fig.(\ref{AngularMSD2})). This quadratic growth with time persists until about $T\sim 1$, which is the typical oscillation time in the system. 
For times $T>>1$, the evolution of the angular MSD slows down and is well described by the relation $\langle\Delta \phi^2(T)\rangle\; = 2 D T$, with $D\approx 0.139$. This linear temporal evolution of the MSD was also observe for other energies, namely $0.27\leq E \leq 0.31$ (see appendix~\ref{app:MSD}). The best fitting curves, for both $T<1$ and $T>>1$, can be found in the appendix appendix~\ref{app:MSD}. We next come to discuss the ballistic initial evolution as well as the late time Brownian-like evolution of the angular MSD in the context of the statistical behavior of enclosed areas, traced by Brownian paths.

\subsubsection{Calculating the MSD through loop statistics of planar Browninan paths}
Recall that the geometric phase can be identified with the area enclosed by the path traced in the planar reparametrized shape space. Consequently, the statistical temporal behavior of the MSD should follow that of the corresponding areas enclosed by the paths. We note that the persisting island of regular trajectories in phase space imply that even for the chaotic trajectories some correlations may persist. This is particularly noticeable in systems of low dimensionality as the one considered here. Higher moment calculations indeed show some deviations from the predicted statistics for a true uncorrelated Brownian path (see appendix~\ref{app:MSD}). Nonetheless, as most of the trajectories for $E=0.3$ show a positive Lyapunov exponent (see figure~\ref{lle}), and a linear variance we next come to approximate them as Wiener processes, bearing in mind that some deviations are to be expected due to underlying long time correlations in the physical trajectories.

Assuming that the phase space trajectories are Brownian curves, then the problem has a well known solution. The probability distribution of the area enclosed by a planar Brownian curve was first studied by Paul L\'{e}vy \cite{levy}, and by now has appeared in the description of a variety of seemingly different physical systems~\cite{maggs1,maggs2,edwards_stat,brereton1987topological,sinha1994brownian}, where the result is often re-derived using the path integral formulation \cite{khandekar1988distribution,brereton1987topological,duplantier1989areas,brown_sphere,sinha1994brownian}. For closed Brownian paths, the probability distribution for the enclosed algebraic area reads~\cite{desbois1992algebraic}:

\begin{equation}
P(A)=\frac{\pi}{T\cosh^2(2\pi A/T)} \;.
\label{PA}
\end{equation}

The probability distribution allows the straightforward calculation of the area second order moment, $\langle A^2\rangle$, which yields a result that is proportional to $T^2$. The direct link between $\langle A^2\rangle$ and the angular MSD would leads us to conclude that the time dependence of the angular MSD should always be proportional to $T^2$, in contradiction with our numerical results. This discrepancy arises from a fundamental difference between the trajectories traced by our system in phase space, and those generated by a na\"ive planar random walk. In the latter, as time progresses loops grow in size indefinitely. This gives rise to L\'{e}vy's original result. However, in our system the motion in phase space is subjected to a restoring potential, and thus at a given energy level only loops of finite maximal distance from the origin can be obtained. We, thus, need to analyse what effects are introduced in the temporal behaviour of $\langle A^2\rangle$ by a confining potential.

The path traced by a particle undergoing Brownian motion produces a Brownian curve. In order to take into account the effect introduced by a confining potential, we consider that this particle moves in the external field of an isotropic quadratic potential. 
While the quadratic potential we use is much simpler than the one that enters the Hamiltonian in eq.(\ref{Hred}), it similarly serves to confine the motion in phase space, and preclude, for a given energy scale, loops of arbitrarily large area.

If instead of moving freely, the particle moves in the external field of a quadratic potential, the probability of tracing a closed loop, that starts and ends at $\mathbf{r}_0$, after a time $\tau=T$ can be written as the path integral (in the large friction limit)~\cite{wiegel1986}:

\begin{equation}
P(\mathbf{r}_0,T) = \int^{\mathbf{r}(T)=\mathbf{r}_0}_{\mathbf{r}(0)=\mathbf{r}_0} d[\mathbf{r}(\tau)] \; e^{- \mathcal{S}} \; ,
\label{brownian_curve}
\end{equation}
where 
\[
\mathcal{S}=\int^T_0 d\tau \Big[ \frac{1}{4 \mathcal{D} } \Big( \frac{d\mathbf{r}}{d\tau} \Big)^2 + V(\mathbf{r}) \Big],
\]
and 
\[
V(\mathbf{r})=\frac{\mathbf{F}^2}{4f^2\mathcal{D}}+\frac{\mathbf{\nabla}\cdot\mathbf{F}}{2f}=-\frac{\alpha}{f}+\frac{\alpha^2}{4\mathcal{D}f^2}(x^2+y^2).
\]
The coefficients $f$ and $\mathcal{D}$ are, respectively, the friction and the diffusion coefficients associated to the Brownian trajectory, and $\mathbf{F}$ is the force field generated by the isotropic quadratic potential, i.e. $\mathbf{F}=-\alpha \; \mathbf{r}$. Thus, $\alpha$ can be seen as the spring constant of the corresponding harmonic potential. 

Our task is to determine the second order moment $\langle A^2\rangle$. In probability theory, the characteristic function, $\mathcal{C}_A(B)$, not only completely defines the corresponding probability distribution function, but also allows for the direct computation of any of the $n$-th order statistical moments through the relation~\cite{probability} 
\[
\langle A^n\rangle=(-i)^n \; \frac{\partial^n}{\partial B^n}\mathcal{C}_A(0),
\]
whenever these are finite. Thus, to obtain $\langle A^2\rangle$, it suffices to derive the characteristic function $\mathcal{C}_A(B)$. By definition $\mathcal{C}_A(B)=\langle e^{iBA}\rangle$, where $A$ is a functional of the underlying Brownian paths and the expectation value is evaluated over Brownian trajectories~\cite{khandekar1988distribution,brown_sphere,desbois1992algebraic,sinha1994brownian}. The algebraic area enclosed by the path depends on the particle's trajectory through the integral:

\begin{equation}
A= \frac{1}{2} \oint (x \; dy \; - \; y \; dx) =  \frac{1}{2} \oint \Big(x \; \frac{dy}{d\tau} \; - \; y \; \frac{dx}{d\tau}\Big) \; d\tau .
\label{area_def}
\end{equation}
The above formulae employs the symmetric gauge for the vector potential associated with the area elements. While for closed trajectories $A$ will necessarily be gauge independent, for open trajectories different gauges can give rise to different results for $A$. However, as we show in appendix~\ref{app:open_paths}, the angular MSD's diffusive behaviour in the limit $T>>1$ is also present for open paths. 

For the time being, $B$ is a real-valued parameter with units of $(\text{length})^{-2}$, whose interpretation will become clear later on. The characteristic function, then, reads as the normalised average of the path functional $e^{iBA}$, over all possible closed Brownian paths defined through eq.(\ref{brownian_curve}), namely:

\begin{equation}
\begin{split}
\mathcal{C}_A(B)&= \frac{ \int d\mathbf{r}_0 \int^{\mathbf{r}(T)=\mathbf{r}_0}_{\mathbf{r}(0)=\mathbf{r}_0} d[\mathbf{r}(\tau)] \; e^{iBA} \; e^{- \mathcal{S}(\tau)}}{\int d\mathbf{r}_0 \int^{\mathbf{r}(T)=\mathbf{r}_0}_{\mathbf{r}(0)=\mathbf{r}_0}  d[\mathbf{r}(\tau)] \; e^{- \mathcal{S}(\tau)}} \\ &= \frac{\int d\mathbf{r}_0 \int^{\mathbf{r}(T)=\mathbf{r}_0}_{\mathbf{r}(0)=\mathbf{r}_0} d[\mathbf{r}(\tau)] \; e^{- \mathcal{\tilde{S}}(\tau)}}{\int d\mathbf{r}_0 \int^{\mathbf{r}(T)=\mathbf{r}_0}_{\mathbf{r}(0)=\mathbf{r}_0} d[\mathbf{r}(\tau)] \; e^{- \mathcal{S}(\tau)}} \; ,
\end{split}
\label{variance_A1}
\end{equation}
with $\mathcal{S}$ and $\mathcal{\tilde{S}}$ given by:
\begin{equation}
 \begin{split}
& \mathcal{S}=\int^T_0 d\tau \mathcal{L} \; , \;\;  \mathcal{L} =  \frac{1}{4\mathcal{D}} \Big( \frac{d\mathbf{r}}{d\tau} \Big)^2 - \frac{\alpha}{f}+\frac{\alpha^2}{4\mathcal{D}f^2}(x^2+y^2) \\ & \mathcal{\tilde{S}}=\int^T_0 d\tau \mathcal{\tilde{L}} \;, \;\;   \mathcal{\tilde{L}} = \mathcal{L} + i\frac{B}{2} \Big( x \; \frac{dy}{d\tau} \; - \; y \; \frac{dx}{d\tau} \Big) \; .
 \end{split}
 \label{lagrangians}
 \end{equation}
The resulting path integrals in eq.(\ref{variance_A1}) share close resemblance with the quantum mechanical path integrals for a harmonically bounded charged particle in a uniform magnetic field (in the numerator), and for the isotropic harmonic oscillator (in the denominator), under the transformation to imaginary time ($t=-i \hbar \beta$). This transformation is the typical approach used in relating the quantum mechanical propagator to the corresponding density matrix in statistical physics~\cite{khandekar,feynman}. The density matrices for a harmonically bounded charged particle, subjected to a uniform magnetic field, and for the isotropic harmonic oscillator are given by: 
\begin{equation}
 \begin{split}
&\rho_{\text{HO}}(\mathbf{r},\mathbf{r}_0,\beta) = \int^{\mathbf{r}(\hbar \beta)=\mathbf{r}}_{\mathbf{r}(0)=\mathbf{r}_0} d[\mathbf{r}(\tau)] \;  e^{-\frac{1}{\hbar} S_{HO} (\mathbf{r}(\tau)) } \;, \\
 & \text{with} \;\; S_{HO} (\mathbf{r}(\tau)) = \int^{\hbar \beta}_0 d\tau \; \bigg \{ \frac{m}{2} \Big( \frac{d\mathbf{r}}{d\tau} \Big)^2 + \frac{m}{2} \Omega^2 \mathbf{r}^2   \bigg\} \; . \\ 
&\rho_{\text{B}}(\mathbf{r},\mathbf{r}_0,\beta) = \int^{\mathbf{r}(\hbar \beta)=\mathbf{r}}_{\mathbf{r}(0)=\mathbf{r}_0} d[\mathbf{r}(\tau)] \;  e^{-\frac{1}{\hbar} S_B (\mathbf{r}(\tau)) } \;, \\
 & \text{with} \;\; S_B (\mathbf{r}(\tau)) = \int^{\hbar \beta}_0 d\tau \; \bigg \{ \frac{m}{2} \Big( \frac{d\mathbf{r}}{d\tau} \Big)^2 + \frac{m}{2} \Omega^2 \mathbf{r}^2 \\& - imw \Big( x \; \frac{dy}{d\tau} \; - \; y \; \frac{dx}{d\tau} \Big)  \bigg\} \; .
\end{split}
 \label{density_matrices}
 \end{equation}
 Neglecting for a moment the constant term $\alpha/f$ in eq.(\ref{lagrangians}), we can recover the remaining terms in eq.(\ref{lagrangians}) from eq.(\ref{density_matrices}) by setting:
 \begin{equation}
     m= \frac{\hbar}{2\mathcal{D}} \; , \; \Omega = \frac{\alpha}{f} \; , \; w =  -B\mathcal{D} \; , \; T = \hbar \beta \; .
 \label{mapping_systems}
 \end{equation}
The term $(-\alpha)/f$ corresponds to adding a constant term to the potential, and consequently its effect is to shift all energy eigenvalues by a constant factor. Therefore, it changes the propagator by multiplying it by an overall constant phase factor. This constant shift in the energies will be the same for both propagators, and thus the extra phase factor drops out when considering their ratio in eq.(\ref{variance_A1}). Furthermore, since $\mathcal{C}_A(B)$ is only evaluated for closed paths, the analogy with the density matrices in eq.(\ref{density_matrices}) tells us that the characteristic function we seek is given by the ratio of the partition functions $Z_B/Z_{HO}$. These are well known~\cite{khandekar,feynman}, which allow us to write:
\begin{equation}
  \begin{split}
     \mathcal{C}_A(B)&= \frac{Z_B(T)}{Z_{HO}(T)}= \frac{\int d\mathbf{r}_0 \; \rho_{\text{B}}(\mathbf{r}_0,\mathbf{r}_0,T)}{\int d\mathbf{r}_0 \; \rho_{\text{HO}}(\mathbf{r}_0,\mathbf{r}_0,T)} \\ &= \frac{\sinh^2\big(\frac{T \Omega}{2}\big)}{\sinh \big(\frac{T w_1(B)}{2} \big) \sinh \big( \frac{T w_2(B)}{2} \big)} \;, 
     \end{split}
     \label{eq:CAB}
\end{equation}  
with $w_j(B)= (-1)^j \;  B\mathcal{D} + \sqrt{B^2 \mathcal{D}^2 + \Omega^2}$ and $j=\{1,2\}$. It then follows that the area second moment $\langle A^2\rangle$ is given by:
\begin{equation}
\langle A^2\rangle=\mathcal{D}^2  \frac{Tf}{\alpha} \Bigg( \frac{\sinh(T\alpha/f)-T\alpha/f}{\cosh(T\alpha/f)-1} \Bigg) \; .
\label{variance_final}
\end{equation}

We have then all the required information to analyse how the $\langle A^2 \rangle$ behaves as a function of time, in particular in the small ($T<<1$) and large time ($T>>1$) limits. For the small time limit, we may expand the hyperbolic functions in Taylor series, and retain only the lowest order terms. This allows us to conclude that
\begin{equation}
\langle A^2\rangle |_{T<<1} \to \frac{1}{3} \mathcal{D}^2 T^2 
\label{small_time_limit}
\end{equation}
which agrees with the quadratic behaviour predicted by L\'{e}vy~\cite{levy}. Indeed, for short times, the particle does not yet feel the confinement that arises from its finite energy. 
However, for longer time scales, eq.(\ref{variance_final}) reduces to
\begin{equation}
\langle A^2\rangle|_{T>>1} \to \frac{\mathcal{D}^2 f}{\alpha} T 
\label{long_time_limit}
\end{equation}
predicting a linear dependence on time. Both of these limits are in qualitative agreement with the numeric results for the MSD in figs.(\ref{AngularMSD1}) and (\ref{AngularMSD2}). Note that while $\langle A^2\rangle$ has units of $(\text{length})^4$, the angular MSD computed in eq.(\ref{AMSD_eq}) is dimensionless. The diffusion coefficient associated with the large time behaviour of $\langle \phi^2(T)\rangle$ in fig.(\ref{AngularMSD1}) should then be interpreted as the effective diffusion coefficient emerging from the underlying Brownian paths, and not the diffusion coefficient $\mathcal{D}$ of the phase space trajectories themselves. Additionally, while for a given energy value the actual phase space trajectories of our system are strictly bound and cannot pass through points arbitrarily far from the origin, the calculated paths that contribute to eq.(\ref{eq:CAB}) have an exponentially small, yet finite, probability of visiting such points. One could argue that instead of using an external quadratic potential, the Brownian particle should be confined to a finite region by applying an infinite potential at the boundaries. Following a perturbative approach the corresponding statistics was calculated
in~\cite{desbois1992algebraic} recovering the linear long time behavior described above. This suggests that the observed long time behaviour of $\langle A^2\rangle$ is indeed a result of paths being traced over a bounded region, rather than an effect induced by the microscopic details of the bounding potential. On the other hand, the short time limit can be understood as the result of the particle having mostly the chance of covering small areas, and therefore the existence of a bounded region does not play a significant role there. For this reason, the statistics of the enclosed area in this limit should not depart too much from L\'{e}vy's result, as is evident by eq.(\ref{small_time_limit}). 

Thus, we expect that, whenever Brownian paths are traced over a bounded region, regardless of the type of confining potential used, $\langle A^2\rangle$ will grow linearly with time, when probed over long enough time scales. Since the angular MSD has a direct relation with the path enclosed area, this result has immediate consequences for the behaviour of the angular MSD, when the Hamiltonian dynamics becomes chaotic.

\section{Summary}
The three degrees of freedom describing the toy model presented here can be separated into  two groups: two shape variables, that uniquely determine the conformation of the system, and one additional variable that determines the system's orientation in space. The conservation of angular momentum yields a non-holonomic constraint for the angular variable, relating its evolution to the history of shape deformations. Such deformation-induced-rotations have been discussed in the context of the cat righting maneuver~\cite{montgomery}, rotations induced by molecular vibrations~\cite{wijn}, and various three mass systems \cite{ori1,montgomery2,molgeophase,square_cat}. The shape-spaces in these systems are defined as quotient spaces identifying all configurations of the systems that differ by rigid motions with each other.  
A cyclic sequence of shape deformations corresponds to a closed path in shape space, and the reorientation associated with it, in the case of two dimensional Hamiltonians takes the form of a geometric phase:
\[
\Delta \phi=\iint f(x^1,x^2)dx^1 dx^2.
\]
It is straightforward to show that one could always find a system of coordinates for shape space $(\tilde{x}^1,\tilde{x}^2)$ on some small neighborhood of any given point with respect to which the integral above assumes the form of a Cartesian area element: \[
\Delta \phi=\iint d\tilde{x}^1 d\tilde{x}^2.
\] 
\footnote{The straightforward example reads $\tilde{x}^2=x^2,\tilde{x}^1=\int f(x^1,x^2)dx^1 $. }
Moser \cite{moser}, demonstrated that this could be done globally. Thus, focusing on the evolution of a single phase, it suffices to study the behavior of planar paths, as done in the present work.

In all the systems described above displaying deformation-induced-rotations, the orientation of the systems can be directly measured by an observer. However, as the orientation depends on the full history of areas enclosed by the path in shape space, it is highly sensitive to correlations in the system's dynamics. Such correlations, for example, were shown to lead to fractional statistics and super-diffusion in the harmonic-three-mass system \cite{ori1,ori2}. Even when these temporal correlations in the system's shape dynamics are lost, as is the case of developed chaos, the statistical moments of the geometric phase carry information on the structure and boundedness of these trajectories, as seen here.  

By interpreting the geometric phase as the path enclosed area, in a suitable shape space, we benefit from L\'{e}vy's seemingly unrelated study of areas traced by closed Brownian paths. We have seen that direct application of L\'{e}vy's area law leads to a squared angular displacement that grows quadratically in time. We identified the source for the seeming ballistic rotation with the unbounded increase of the typical area element with time. This behaviour is only followed in the initial stages of the evolution in our system. After a finite time we observe that the squared angular displacement follows a linear scaling with time associated with the saturation of the magnitude of the typical area element to a finite average value. This should always be the case for systems whose phase space trajectories explore only a bounded region in phase space. In our case the restoring elastic potential bounds the possible strains for any given total energy. We approximated this confinement by considering a random walker subjected to a harmonic potential centered at the origin. This renders far excursions in phase space exponentially improbable and restores the diffusive scaling for the evolution of the orientation. Replacing the restoring harmonic elastic potential with an infinite potential well also recovered the diffusive behavior~\cite{desbois1992algebraic}. We expect the resulting statistics to not depend on the details for the confining potential provided that it is relatively isotropic, convex and strong enough to preclude arbitrary far excursions form the origin. We believe that the mechanism presented here may be responsible for the diffusive-like statistics of non-holonomically constrained variables in a wide variety of chaotic Hamiltonian systems. 

\begin{acknowledgments}
This work was supported by the Minerva Foundation with funding from the Federal German Ministry for Education and Research, and by the Israel Science Foundation Grant No. 1444/21.
\end{acknowledgments}

\section*{Author Declarations}
\subsection*{Conflict of Interest}
The authors have no conflicts to disclose.
\subsection*{Author Contributions}
\textbf{Ana Silva:} Conceptualization (equal); Methodology (lead); Formal analysis (lead); Writing – original draft (equal). \textbf{Efi Efrati:} Conceptualization (equal); Supervision (lead); Methodology (supporting); Formal analysis (supporting); Writing – original draft (equal).

\section*{Data Availability Statement}
The data that support the findings of this study are available from the corresponding author upon reasonable request.

\appendix

\section{Largest Lyapunov exponent} \label{app:lle}
\begin{figure}
 \includegraphics[width=1.0\columnwidth]{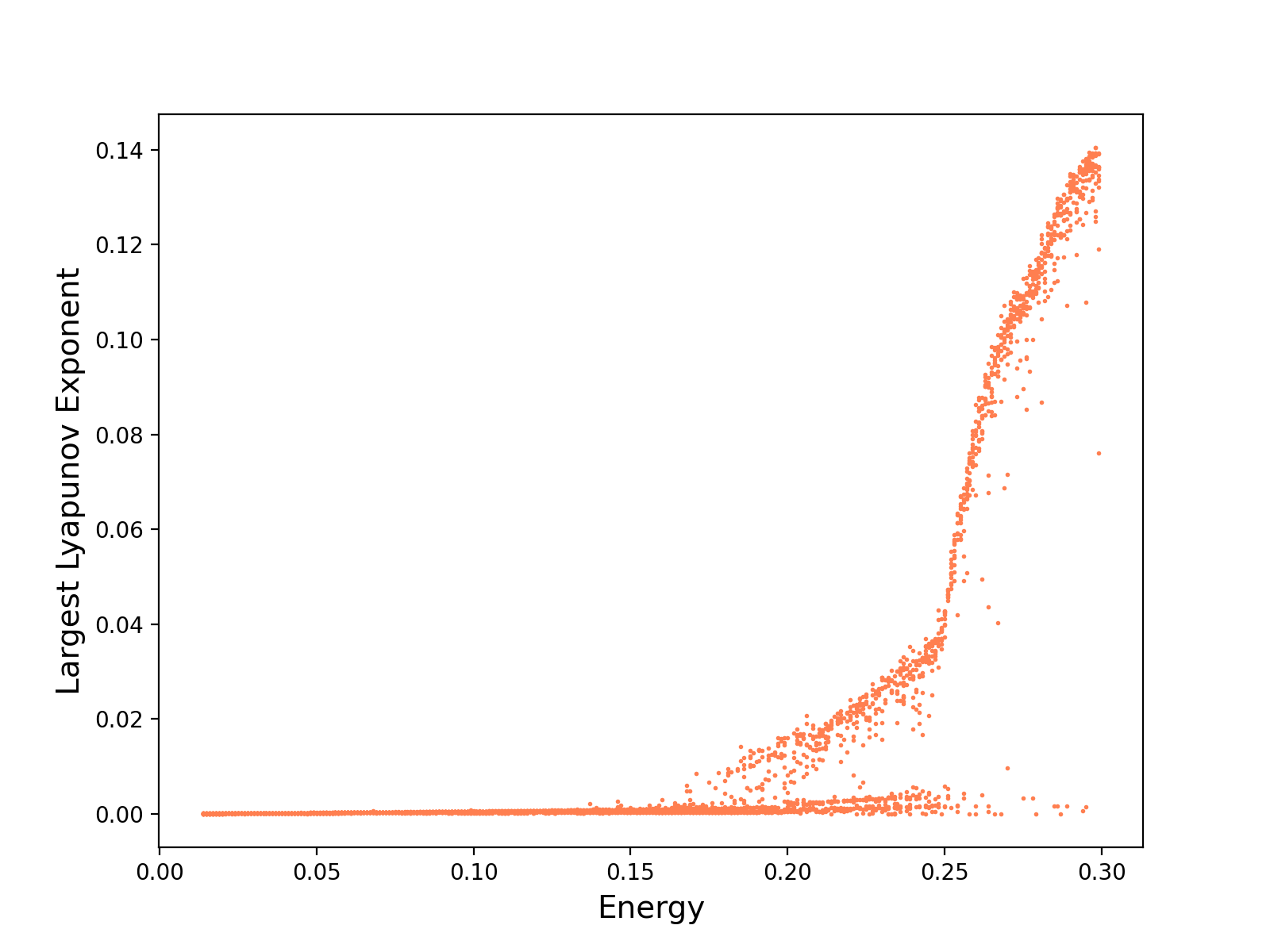}
\caption{Estimate of the largest Lyapunov exponent for the system described in the main text, as a function of its energy. The data points where obtained by running the code twelve times, with random initial conditions that match the corresponding energy, and by tacking the energy to be incremented in steps $\Delta E=0.001$. At $E=0.3$, the largest Lyapunov exponent assumes its largest positive value, within this energy range.\label{lle}}
\end{figure}

The dynamics of the triangular shaped system with two identical masses $m$ was described via the reduced Hamiltonian in eq.(\ref{Hred}). To study the qualitative behaviour of the resulting trajectories, we obtained the Poincar\'{e} sections for different energy values ($E$) of the system, and for the set of fixed parameters $m=1$, $k_1=k_2=1$ and $b=L=1$. The corresponding Poincar\'{e} section at $E=0.3$ showed a collection of randomly scattered points covering a finite area in the plane $(a,p_a)$, a result consistent with the presence of chaotic dynamics. This supported our claim that our system exhibits both regular and chaotic dynamical regimes. To further corroborate this observation, we numerically obtained the largest Lyapunov exponent as a function of the system's energy $E$ (see fig.(\ref{lle})). The energy domain was restricted to the interval $E \in [0.014,0.3]$, to coincide with the energy range at which the Poincar\'{e} sections were studied.\\
The Lyapunov exponents measure the rate at which initially close-by trajectories diverge from one another, along different directions, as time progresses~\cite{skokos2010lyapunov,lowenstein2012essentials}. The symplectic structure of autonomous Hamiltonian systems gives rise to a set of Lyapunov exponents that come in pairs of opposite signs, and thus their sum is constrained to be zero~\cite{skokos2010lyapunov}.  For regular motion, initially nearby trajectories remain nearby, and so all the Lyapunov exponents are zero. In contrast, chaotic motion gives rise to at least one direction for which initially close-by trajectories will exponentially diverge in time, implying at least one positive Lyapunov exponent. Thus, a commonly used method to evaluate the presence of chaotic motion is to numerically probe when the largest Lyapunov exponent is positive~\cite{book_lyapunov_1,skokos2010lyapunov}. 
In fig.(\ref{lle}), we see that 
starting around $E\sim 0.15$ the largest Lyapunov exponent becomes positive. Additionally, its value continuously increases in the energy range examined. In particular, the exponent reads $0.14$ at the energy $E=0.3$ for which the chaotic Poincar\'{e} section was obtained. 

The numerical computation of the largest Lyapunov exponent was based on the standard method described in~\citep{book_lyapunov_1,skokos2010lyapunov,contopoulos1978number}, which requires solving, simultaneously, both the equations of motion and the variational equations (i.e. the linearised versions of the equations of motion). The resulting system of coupled equations was solved using a second order symplectic integrator~\cite{symplectic_int}.\\

In fig.(\ref{lle}), the energy is varied over the interval $E \in [0.014,0.3]$, in steps $\Delta E=0.001$. For each energy, the algorithm determining the largest Lyapunov exponent picks a collection of twelve random initial conditions for the equations of motion, constrained to give rise to the desired energy. 
Fig.(\ref{lle}) shows a clear trend: in the low energy regime, the largest Lyapunov exponent tends to zero, implying regular dynamics, while for the larger energy values the same Lyapunov exponent is clearly positive, indicative of chaotic trajectories. The behaviour of the largest Lyapunov exponent in fig.(\ref{lle}) agrees with the expected behavior for mixed chaotic systems, and is in agreement with the characteristic behavior that emerged from the Poincar\'{e} sections for the selected few energy values examined. 

\begin{figure}[h]
 \includegraphics[width=1.0\columnwidth]{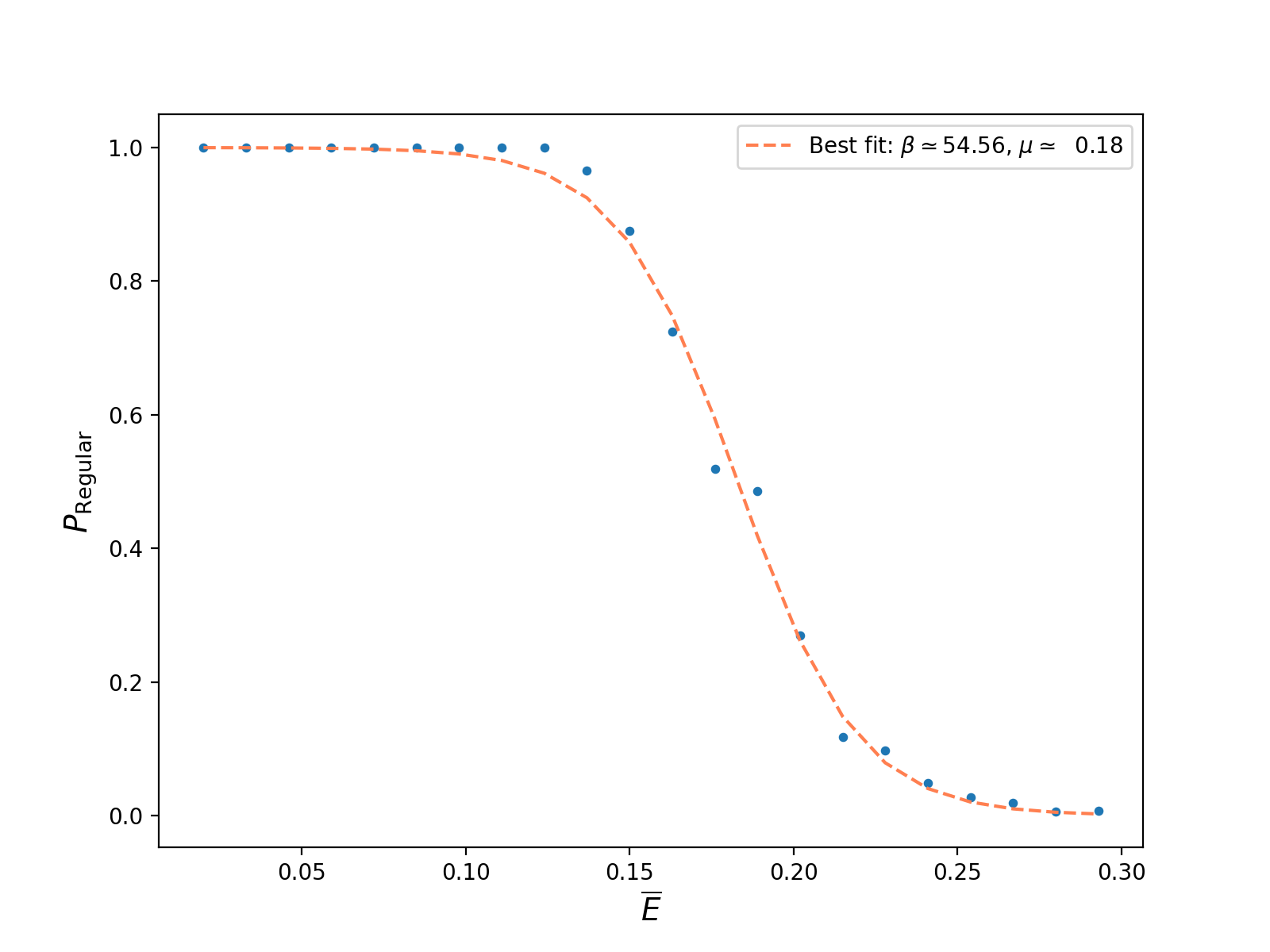}
\caption{Probability of finding a regular trajectory ($P_{\text{Regular}}$) in a given energy shell ($\Delta E_n$). The horizontal axis $\overline{E}$ represents the average energy value associated to each energy shell $\Delta E_n$. Note that there are in total $22$ distinct $\Delta E_n$. An energy shell is defined as the energy interval: $\Delta E_n=[E_0+ \Delta E (n-1) i_{max}$, $E_0 + \Delta E (n \; i_{max} -1)]$, with $E_0=0.014$, $\Delta E=0.001$ and $i_{max}=13$. In the figure, the blue dots represent our estimate of $P_{\text{Regular}}$. The orange curve is the logistic function $(1+e^{\beta (\bar{E}-\mu)})^{-1}$, evaluated for the best fit parameters: $\beta \simeq 54.56$ and $\mu \simeq 0.18$. \label{frac_lle}}
\end{figure}

In the vicinity of $E\sim 0.15$, where the first positive Luyapunov exponents appear, most of the trajectories examined exhibited regular behavior. As the energy increased the fraction of regular trajectories decreased, until for high energy almost all initial conditions yield chaotic trajectories. To further examine this transition, we computed the probability of finding a regular trajectory for each energy value (see fig.(\ref{frac_lle})). For this aim, we divided the energy interval $E \in [0.014,0.3]$ into $N=22$ different subintervals, which we call energy shells. Each energy shell, $\Delta E_n$, is defined as the energy interval $\Delta E_n=[E_0+ \Delta E (n-1) i_{max}, E_0 + \Delta E (n \; i_{max} -1)]$, with $E_0=0.014$, $\Delta E=0.001$, $n=\{1,2,...,N\}$ and $i_{max}=13$ (the total number distinct energy values in each $\Delta E_n$). For a given $\Delta E_n$, we compute what proportion of the corresponding trajectories can be classified as regular. Each energy shell includes $13$ distinct energy values, for each of which $12$ initial conditions were examined. 
Fig.(\ref{frac_lle}) displays the fraction of the $12\times 13$ data points for each energy shell that produce a vanishing largest Lyapunov exponent, indicative of regular dynamics.

The classification of the dynamics as regular occurs when the largest Lyapunov exponent is effectively zero. More precisely, numerical evaluation of largest Lyapunov exponent provides only an estimate, and even the numerical computation of the Lyapunov exponents for genuine integrable systems can render non-zero values, albeit always small~\cite{numericalLyapunov}. Thus, we need to set an upper limit to the value of the largest Lyapunov exponent, above which we no longer deem the dynamics regular. All values bellow this upper limit are considered effectively zero, and the dynamics is said to be regular. Hence, whenever the Lyapunov exponent was $< 0.001$, the dynamics was classified as regular. Even though the threshold value of $0.001$ is somewhat arbitrary, we note that it is still reasonably small and it is $2$ orders of magnitude smaller, as compared to the maximum value attained by the largest Lyapunov exponent over the energy range $E \in [0.014,0.3]$. 

In fig.(\ref{frac_lle}), we show the probability of finding a regular trajectory for each energy shell $\Delta E_n$ as a function of the average energy in the shell. The blue dots represent the obtained probability, while the orange curve is given by the logistic function $(1+e^{\beta (\Delta \bar{E}-\mu)})^{-1}$ with $\beta\approx 57.56$ and $\mu \approx 0.18$. The value of the logistic function parameters, $\beta$ and $\mu$, was fixed at the best fit found for the numerical data (blue dots). The fit was attained using Python's open-source scientific computing library Scipy, namely the built-in function curve$\_$fit. The parameter $\beta$ controls the steepness of the orange curve: the higher it is, the faster the probability decays towards zero. On the other hand, the value of $\mu$ defines for which energy shell $\Delta E_n$ the probability $P_{\text{Regular}}$ becomes exactly $1/2$. The obtained value for $\mu$ indicates that $P_{\text{Regular}}$ would be $1/2$ in an energy shell, whose average energy value would be around $0.18$. Indeed, in fig.(\ref{lle}) the rise of the largest Lyapunov exponent starts to become apparent around $E \in ]0.15;0.2[$, in agreement with the previous prediction set by the value of $\mu$.

Inspection of fig.(\ref{frac_lle}) reveals that for low energy almost all initial conditions result in regular trajectories, in agreement with the identification of the non-linear term in the potential energy as the cause for the loss of integrability. For small enough energies, this term can be truncated to a finite order in its Taylor series. The resulting truncated Hamiltonian is integrable, but as the energy increases more terms of the expansion need to be included. Eventually, the non-linearities pull the Hamiltonian further and further away from its integrable low energy limit. In this sense, we may understand these results in the framework of KAM's theorem, with the energy functioning, effectively, as the perturbation parameter~\cite{ori1,ori2}. As the dynamical system is three dimensional the tori separate space, trapping between them the chaotic regions. As a result, Arnold diffusion is suppressed and the deviation of the chaotic trajectories from their integrable counterparts is bounded. Hence, larger chaotic domains in phase space are expected to take place only for relatively high energy values.     

\section{Fitting the numerical angular MSD to a power law} \label{app:MSD}
\begin{figure}
 \includegraphics[width=0.95\columnwidth]{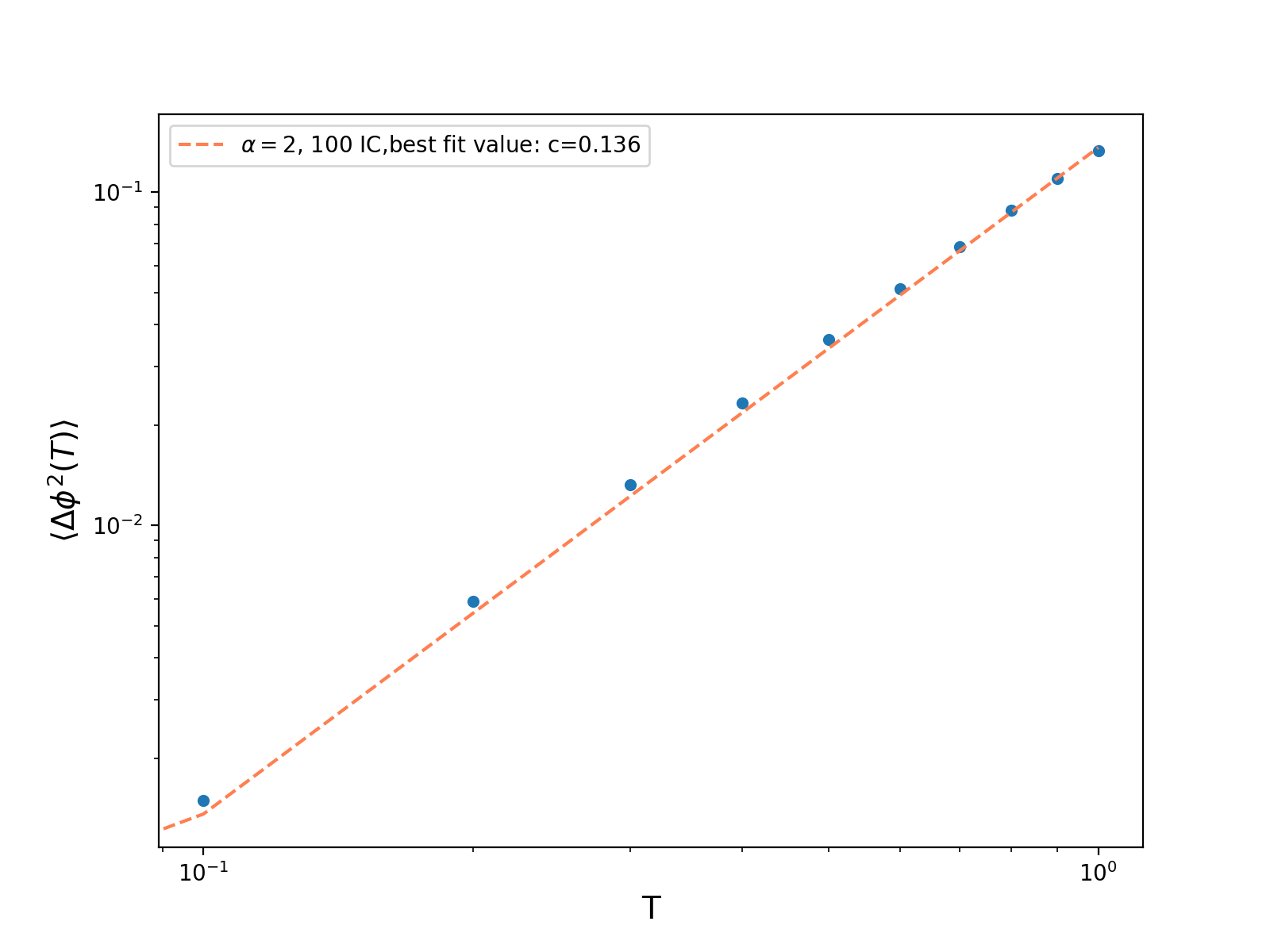}
\caption{When $T$ is small enough, the MSD (blue dots) follows well the power law $\langle\Delta \phi^2(T)\rangle\; \propto  T^{\alpha}$, with $\alpha=2$. The orange line is produced by fitting the low time limit, $T \leq 1$, to the line $c\; T^2$. The best fit was obtained with $c \approx 0.136$. Each data point (blue dotes) in the figure was obtained using an ensemble of 100 initial conditions (IC), as described in the main text.} \label{AngularMSD2}
\end{figure}

\begin{figure}
 \includegraphics[width=0.95\columnwidth]{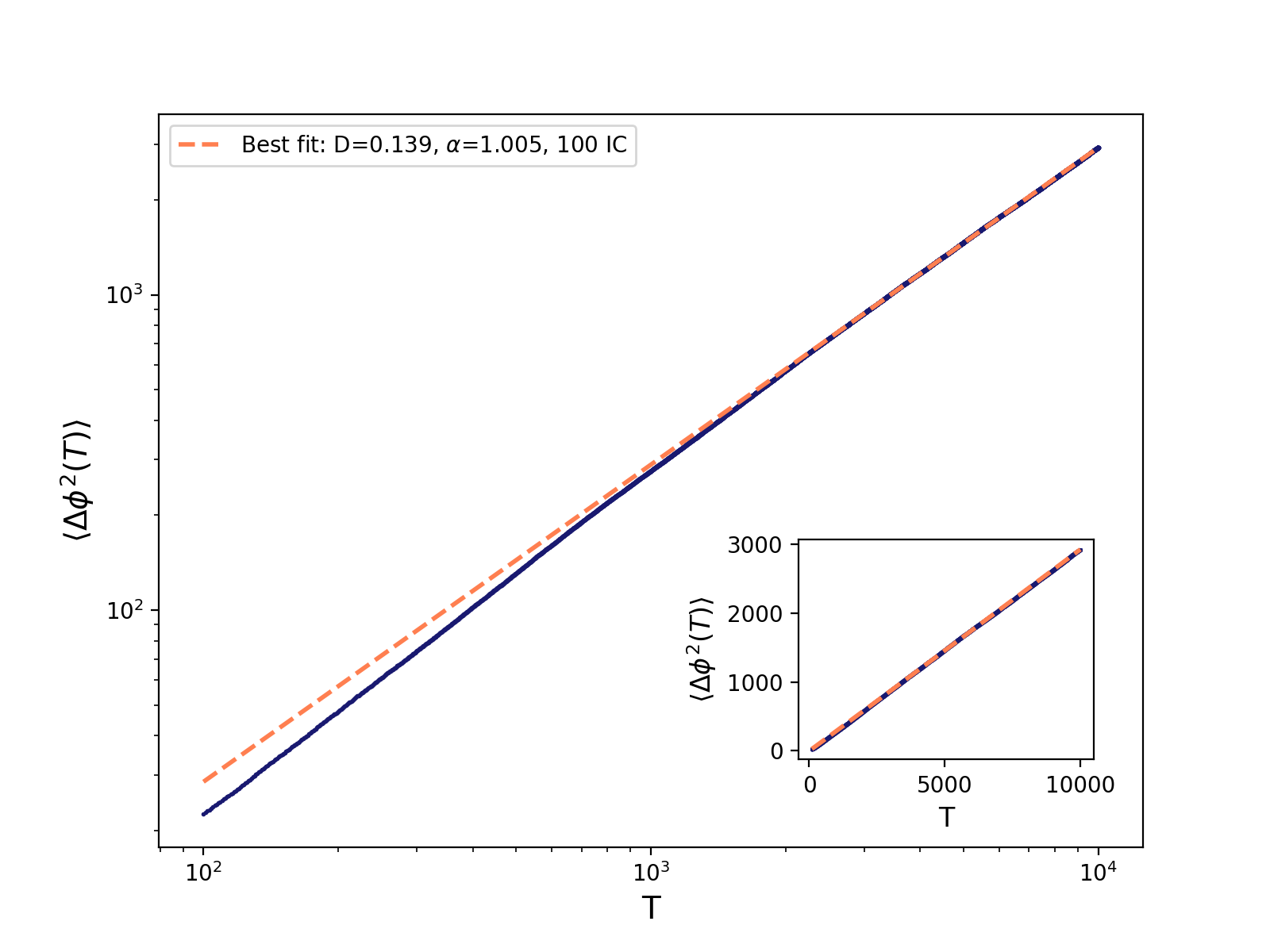}
\caption{The orange line is produced by fitting the angular MSD (blue dots) in the large time limit, $T\in [10,10^4]$, to the power law $2D\; T^\alpha$. The best fit was obtained with $D \approx 0.139$ and $\alpha \approx 1$. Each data point (blue dotes) in the figure was obtained using an ensemble of 100 initial conditions (IC), as described in the main text. In the main figure, both axes are show in logarithmic scale. The inset shows the angular MSD versus best fit, with both axes in linear scale. \label{AngularMSD3}}
\end{figure}

\begin{figure}
\includegraphics[width=0.95\columnwidth]{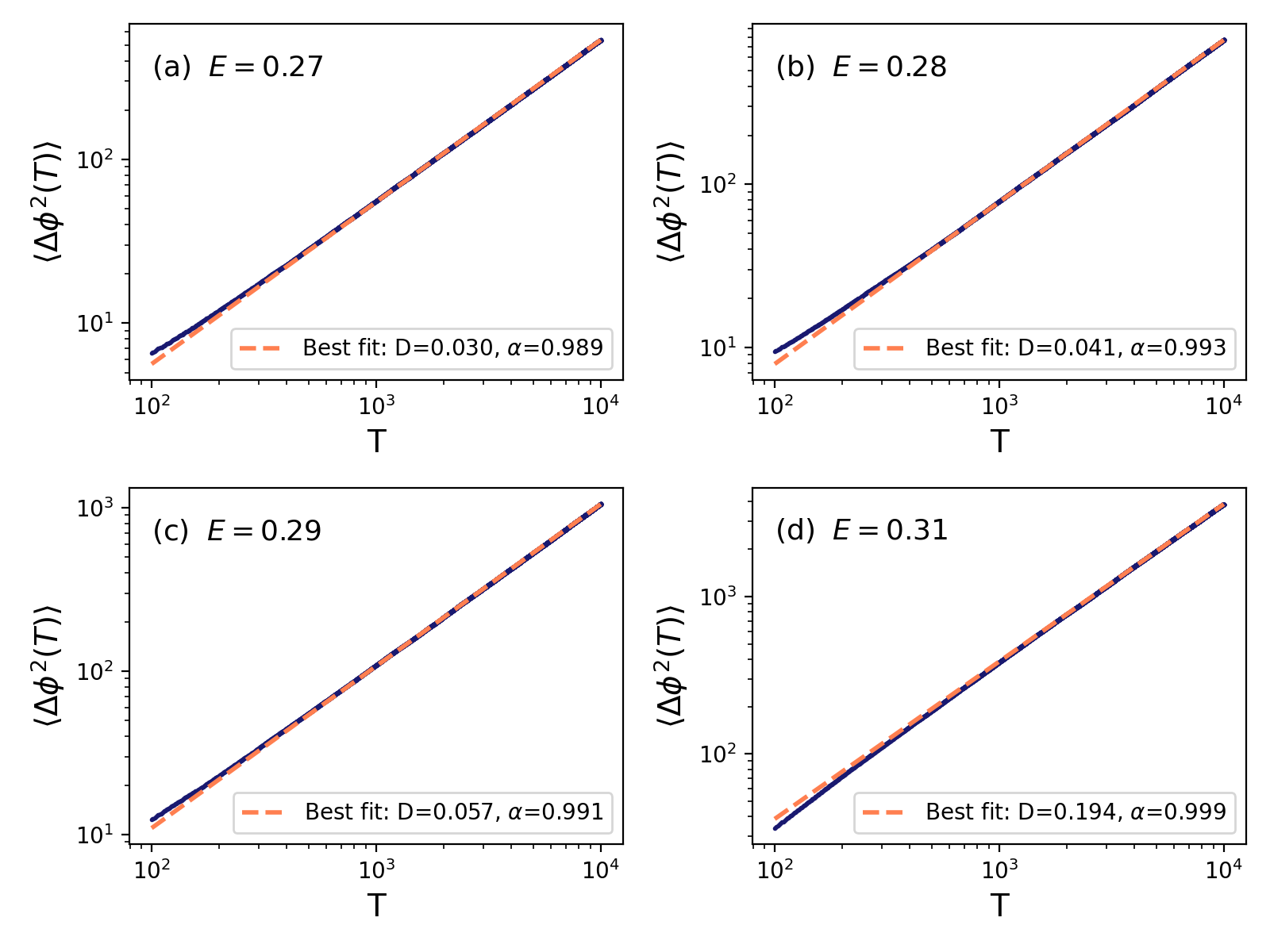}
\caption{The orange line in each figure is produced by fitting the angular MSD (blue dots) in the large time limit, $T\in [10,10^4]$, to the power law $2D\; T^\alpha$.  Each data point (blue dotes) in the figure was obtained using an ensemble of 100 initial conditions (IC), as described in the main text. The fits yield an exponent $\alpha$ close to one, consistent with the typical power law observed for regular diffusion. In all figures, both axes are in logarithmic scale. \label{AngularMSD_4Energies}}
\end{figure}

\begin{figure}
\includegraphics[width=0.95\columnwidth]{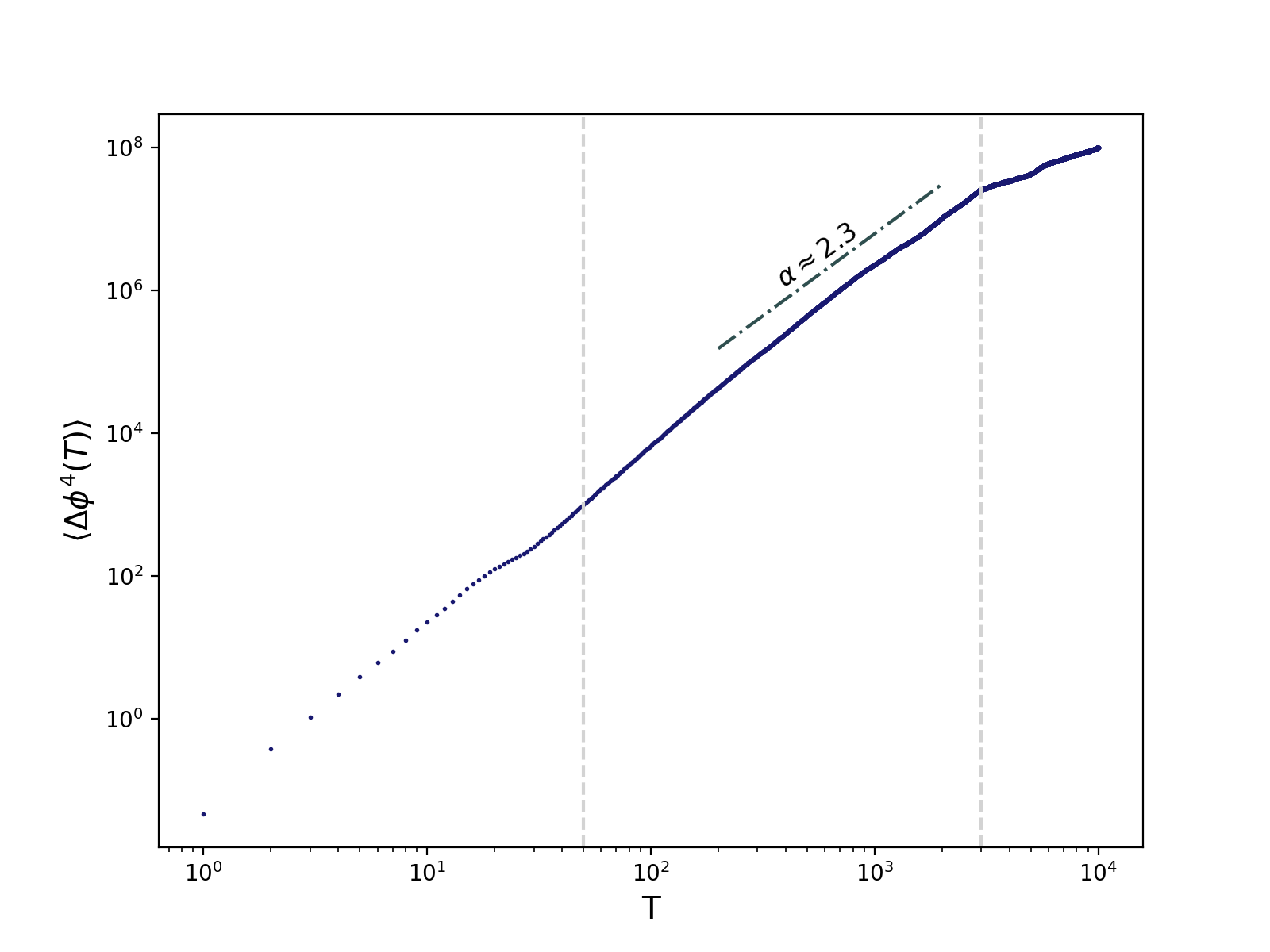}
\caption{The resulting higher order moment $\langle\Delta \phi^4(T)\rangle$ for E=0.3, as a function of time $T$. The dashed gray lines correspond to the values $T=50$ and $T=3000$. Only within this range do we get a closer match with the resulting forth order moment for Brownian paths, but even then the discrepancy is still evident by the mismatch in $\alpha$.\label{mom4}}
\end{figure}

The angular MSD exhibits two distinct power-laws, namely a ballistic trend for $T<1$ and a diffusive behaviour for $T>>1$. These trends were obtained by fitting the numerical data to the corresponding power laws:
\begin{equation}
\begin{split}
   & \langle \phi^2(T) \rangle = c \; T^2 \; , \; \text{for} \; T<1 \; \\
   & \langle \phi^2(T) \rangle = 2D \; T^\alpha \; , \; \text{for} \; T>>1 \; .
    \end{split}
\end{equation}
Figures~\ref{AngularMSD2} and~\ref{AngularMSD3} show the power laws attained for the best fit parameters versus the numerical evaluated angular MSD. The best fits were obtained using Python's open-source scientific computing library Scipy, more precisely its curve$\_$fit function.   
The same linear trend was observed for other energies above $E=0.25$, for $T>10^2$. In fig.~\ref{AngularMSD_4Energies}, we show the angular MSD and the corresponding best fit for energies $E\in\{0.27,0.28,0.29,0.31\}$, for T well above the typical oscillation time $(T \approx 1)$.
Despite the alignment between the growing chaotic sea observed for the corresponding Poincar\'{e} sections in fig.~\ref{PS}, alongside with the growth trend of the largest Lyapunov exponent (fig.~\ref{lle}) for energies above $E=0.25$ and up until around $E=0.3$, as well as the diffusive trend of the angular MSD for $T>>1$, the system still retains non-trivial correlations. This can be made more transparent by evaluating higher order moments. Here, we restricted ourselves to the computation of the fourth order moment $\langle \phi^4(T) \rangle$, for $E=0.3$. The resulting outcome can be found in fig.~\ref{mom4}. Its computation follows the same strategy as the one introduced for the angular MSD in the main text, i.e.
\begin{equation}
    \langle\Delta \phi^4(T)\rangle=\frac{1}{p v} \sum^{p}_{i=1} \sum^{v}_{m=1} \Big( \phi_{i} \big(T+(m-1)\tau \big)-\phi_{i} \big((m-1)\tau \big) \Big)^4 \; ,
\end{equation}
with $p=100$ denoting the total number of equal energy initial conditions used to produce the trajectories. This set of initial conditions are evolved in time, up until $t_{\text{max}}=10^7$. Each resulting trajectory is then sliced into smaller temporal blocks of length $\tau=10^4$. The total number of blocks is given by $v=\frac{t_{\text{max}}}{\tau}=10^3$. Treating each block as a new trajectory, the final ensemble consists of a total of $pv=10^5$ trajectories. 
If the underlying paths would exactly match the behaviour of Brownian curves, then we would expect the probability distribution of enclosed areas to converge to Gaussian statistics in the large time limit~\cite{desbois1992algebraic}. Under this assumption, fitting the resulting $\langle\Delta \phi^4(T)\rangle$ with the power law $c \;T^{\alpha}$ would give rise to $\langle\Delta \phi^4(T)\rangle \to 3(2D T)^2 $. Equivalently, from eq.(\ref{eq:CAB}), the computation of the forth order moment, in the large time limit, yields:
\begin{equation}
    \langle A^4\rangle \underset{T>>1}{\to} 3 \frac{\mathcal{D}^4}{\Omega^2} T^2 = 3  \big( \langle A^2\rangle_{T>>1} \big)^2 \; .
\end{equation}
In fig.~\ref{mom4}, only the region where $T\in [50, 3000]$ comes closer to this expectation, with $c \approx 12D^2-0.0319$, but $\alpha\approx 2.3$, instead of $\alpha=2$. The discrepancy between $\langle\Delta \phi^4(T)\rangle$ and its Brownian counterpart implies that the underlying paths do not map entirely to Brownian curves and instead retain some correlations.

\section{Calculating the behaviour of $\langle A^2\rangle$ for open paths} \label{app:open_paths}
The geometric phase $\Delta \phi$ is a gauge independent quantity only for closed paths in shape space. From a practical stand point, however, numerical evaluation of the angular MSD, $\langle \Delta \phi^2(T) \rangle$, is less time consuming if the average can be preformed without requiring specific demands on the paths. 
In the numeric evaluation of the angular MSD, $\Delta \phi$ will always be zero at $T=0$.
While the MSD for open paths is independent of the initial orientation $\phi(0)$ of the system, we observe a weak dependence on the initial condition in shape space. This is because the confining potential breaks the translational symmetry of the shape space trajectory as a random motion in free space. Consequently, if the system is initiated at an arbitrary point in phase space (a distorted conformation containing a lot of strain), a fast motion, down the gradient of the confining potential leads to an additive term in the MSD. We can simply avoid this sensitivity by considering in our calculations paths that start at the origin, which corresponds to the undistorted shape. Note that in the spring mass system the total energy of the system strongly limits the possible strains at the initial condition and thus this issue does not arise. 
Thus, in our calculations we now examine what happens to the area's second order moment $\langle A^2\rangle$ for paths starting at the origin of a reference frame, but ending at possible distinct points in space. Thus, allowing the paths that contribute to the area statistics to also form open paths. 

We again consider that these paths are traced by a Brownian particle that moves in the external field of a quadratic potential. It is known that for spatially unbounded Brownian paths, the behavior of $\langle A^2\rangle$ follows the same $T^2$ scaling predict by L\'evy for both closed and open paths, with the origin fixed at the starting point. Their temporal power laws will yet be weighted by different multiplicative pre-factors, even if they have the same temporal scaling~\cite{desbois2015}. In what follows, we verify that the same temporal scaling observed for $\langle A^2\rangle$ in the main text remains unchanged, even when the closure of the paths is not enforced. 

In order to evaluate $\langle A^2\rangle$ for open paths, it suffices to construct the characteristic function $\mathcal{C}_A (B)= \langle e^{iBA} \rangle$, where the signed area $A$ is a functional of the Brownian paths~\cite{duplantier1989areas,brown_sphere,sinha1994brownian,mashkevich2009area,desbois2015}. Therefore, the expectation value $\langle e^{iBA} \rangle$ needs to be evaluated as an average over all Brownian paths:
\begin{equation}
\mathcal{C}_A(B) = \frac{\int d\mathbf{r} \; e^{\frac{\beta}{2} \int^{\mathbf{r}}_{\mathbf{0}} \mathbf{F}(\mathbf{r''})\mathbf{.}d\mathbf{r''}}  \int^{\mathbf{r}(T)=\mathbf{r}}_{\mathbf{r}(0)=\mathbf{0}} d[\mathbf{r}(\tau)] \; e^{iBA} e^{-\mathcal{S}} }{\int d\mathbf{r} \; e^{\frac{\beta}{2} \int^{\mathbf{r}}_{\mathbf{0}} \mathbf{F}(\mathbf{r''})\mathbf{.}d\mathbf{r''}  }   \int^{\mathbf{r}(T)=\mathbf{r}}_{\mathbf{r}(0)=\mathbf{0}} d[\mathbf{r}(\tau)] \; e^{-\mathcal{S}}} \; ,
\label{characteristic_function_open1}
\end{equation}
 where as before
\begin{equation}
\begin{split}
\mathcal{S}&= \int^T_0 d\tau \; \mathcal{L}_0(\dot{\mathbf{r}},\mathbf{r},\tau)=\\ 
&=\int^T_0 d\tau \Bigg( \frac{\dot{\mathbf{r}}^2}{4\mathcal{D}}  - \frac{\alpha}{f} + \frac{\alpha^2}{4\mathcal{D}f^2} (x^2+y^2)  \Bigg) \;,
\end{split}
\label{action}
\end{equation}
with $f$ and $\mathcal{D}$ the friction and diffusion coefficients associated to the Brownian trajectories and with $\alpha$ a positive constant. We note that in contrast with the main text, the propagators now exhibit an extra phase factor proportional to the inverse temperature $\beta$, namely $e^{\frac{\beta}{2} \int^{\mathbf{r}}_{\mathbf{r'}} \mathbf{F}(\mathbf{r''}) \cdot d\mathbf{r''}}$~\cite{wiegel1986}. For closed paths, this phase factor evaluates to zero, since the harmonic force under which the Brownian motion occurs is a conservative force. Its presence remains for open paths. 
As outlined in the main text, a closely related set of propagators are the propagators representing the density matrix of a harmonically bounded charge particle, subjected to a constant magnetic field perpendicular to the plane of motion, $\rho_{\text{B}}$, and the density matrix for the two-dimensional isotropic harmonic oscillator, $\rho_{\text{HO}}$,~\cite{}:
\begin{equation}
 \begin{split}
&\rho_{\text{HO}}(\mathbf{r},\mathbf{r}_0,\beta) = \int^{\mathbf{r}(\hbar \beta)=\mathbf{r}}_{\mathbf{r}(0)=\mathbf{r}_0} d[\mathbf{r}(\tau)] \;  e^{-\frac{1}{\hbar} S_{HO} (\mathbf{r}(\tau)) } \;, \\
 & \text{with} \;\; S_{HO} (\mathbf{r}(\tau)) = \int^{\hbar \beta}_0 d\tau \; \bigg \{ \frac{m}{2} \Big( \frac{d\mathbf{r}}{d\tau} \Big)^2 + \frac{m}{2} \Omega^2 \mathbf{r}^2   \bigg\} \; . \\ 
&\rho_{\text{B}}(\mathbf{r},\mathbf{r}_0,\beta) = \int^{\mathbf{r}(\hbar \beta)=\mathbf{r}}_{\mathbf{r}(0)=\mathbf{r}_0} d[\mathbf{r}(\tau)] \;  e^{-\frac{1}{\hbar} S_B (\mathbf{r}(\tau)) } \;, \\
 & \text{with} \;\; S_B (\mathbf{r}(\tau)) = \int^{\hbar \beta}_0 d\tau \; \bigg \{ \frac{m}{2} \Big( \frac{d\mathbf{r}}{d\tau} \Big)^2 + \frac{m}{2} \Omega^2 \mathbf{r}^2 \\& - imw \Big( x \; \frac{dy}{d\tau} \; - \; y \; \frac{dx}{d\tau} \Big)  \bigg\} \; ,
\end{split}
 \label{density_matrices_sm1}
 \end{equation}
which, using polar coordinates, evaluate to~\cite{khandekar}:
\begin{equation}
 \begin{split}
 & \rho_{\text{B}}(\mathbf{r},\mathbf{r}_0,\beta) = \frac{ m\tilde{\Omega}}{2\pi \hbar \sinh(\tilde{\Omega} \hbar \beta)} 
 \exp \big( -g (\mathbf{r},\mathbf{r}_0)\big) \;, \\
&\rho_{\text{HO}}(\mathbf{r},\mathbf{r}_0,\beta) = \lim_{w=0} \rho_{\text{B}}(\mathbf{r},\mathbf{r}_0,\beta) \;, \\
&\text{with } g(\mathbf{r},\mathbf{r}_0) =  \frac{ m \tilde{\Omega}}{2 \hbar \sinh(\tilde{\Omega} \hbar \beta)} \Big( (r^2 +r^2_0) \cosh(\hbar \beta \tilde{\Omega}) \\ &-2 r_0r \cos(\theta-\theta_0+ i\hbar\beta w) \Big)  \; \text{and } \tilde{\Omega} = \sqrt{\Omega^2 + w^2} \; .
 \label{density_matrices_sm2}
 \end{split}
  \end{equation}
Maintaining the same gauge choice for the algebraic area as in the case of closed paths, the term $e^{iBA}$ in eq.(\ref{characteristic_function_open1}) reads:
\begin{equation}
  e^{iBA} =   \exp \Big[ i\frac{B}{2}\int^T_0 d\tau \; \Big( x \frac{dy}{d\tau} - y \frac{dx}{d\tau} \Big) \Big] \; . \nonumber
\end{equation}

We may then relate the path integrals in eq.(\ref{density_matrices_sm1}) with those in eq.(\ref{characteristic_function_open1}) by setting, as in the main text,
\begin{equation}
    m = \frac{\hbar}{2\mathcal{D}} \; , \; \hbar\beta =T \; , \; \Omega = \frac{\alpha}{f} \; , \; w = -B\mathcal{D} \; .
\end{equation}
The characteristic function can then be expressed as:
\begin{equation}
\begin{split}
\mathcal{C}_A(B) & =\frac{\int^{+\infty}_0 dr \; r\; e^{-\frac{r^2}{4} \Big(\alpha \beta + \frac{\tilde{\Omega}}{\mathcal{D}} \coth(\tilde{\Omega} T)\Big) }}{\int^{+\infty}_0 dr \; r\; e^{-\frac{r^2}{4} \Big(\alpha \beta + \frac{\Omega}{\mathcal{D}} \coth(\Omega T)\Big) }} \\ 
&= \tilde{\Omega} \; \frac{\sinh(\Omega T) + \cosh(\Omega T)}{\Omega \sinh(\tilde{\Omega} T) + \tilde{\Omega} \cosh(\tilde{\Omega} T)} \; ,
\end{split}
\label{characteristic_function_open2}
\end{equation}
where we have made use of eq.(\ref{density_matrices_sm2}), and set $\mathbf{r}_0=\mathbf{0}$. Additionally, we have used the Einstein relation to express the inverse temperature $\beta$ in terms of the diffusion and friction coefficients associated to the Brownian trajectories~\cite{wiegel1986}. The second order statistical moment $\langle A^2 \rangle$ can now be evaluated directly from the characteristic function:
\begin{equation}
\begin{split}
    \langle A^2 \rangle &= - \lim_{B=0} \frac{\partial^2 \mathcal{C}_A(B)}{\partial B^2} \\ 
    &= \frac{\mathcal{D}^2}{\Omega^2} \; \big( T \Omega - e^{-T\Omega} \sinh(T \Omega) \big) \; .
    \end{split}
\end{equation}
At small time scales, $\langle A^2 \rangle$ may be expanded in Taylor series, which to lowest order yields a quadratic scaling with time:
\begin{equation}
    \langle A^2 \rangle \underset{T<<1}{\to} \mathcal{D}^2 T^2 \; ,
\end{equation}
At large time scales, $\langle A^2 \rangle$ recovers the linear scaling in time:
\begin{equation}
    \langle A^2 \rangle \underset{T>>1}{\to} \frac{\mathcal{D}^2}{\Omega} T \; .
\end{equation}
Thus, we see that the quadratic behaviour predicted for $ \langle A^2 \rangle$, in the short time limit, and the linear behaviour, in the long time limit, are not exclusive features of closed paths. They also emerge when the path closure constraint is relaxed.

\section{The reduced Hamiltonian in ($\tilde{a}$,$\theta$) coordinates}

We recall that relating the change in orientation with the area enclosed by trajectories, i.e. those that start and end at the same triangular shape, relied on a change of coordinates to ($\tilde{a}$,$\theta$). We now re-write the reduced Hamiltonian in the main text (eq.(\ref{Hred})) in terms of  the new set ($\tilde{a}$,$\theta$). For the sake of clarity, we start from the Lagrangian defined in eq.(\ref{eq:lagrangian}), namely:
\begin{equation*}
\mathcal{L}= \frac{m}{2} \Big( \dot{a}^2 + (a+L)^2 \dot{\phi}^2 + b^2 (\dot{\theta}+\dot{\phi})^2 \Big) + V_{12}(a) + V_{23}(a,\theta) \; ,
\end{equation*}

with the spring potentials $V_{12}(a)$ and $V_{23}(a,\theta)$ given by

\begin{equation*}
\begin{split}
&V_{23}(a,\theta)=\frac{k_2}{2} \Big( \sqrt{b^2+(L+a)^2 -2b(L+a)\cos(\theta)} - L \Big)^2, \\
&V_{12}(a) = \frac{k_1}{2} \Big(|a +L|-L\Big)^2 \;. \end{split}
\label{spring_potentials2}
\end{equation*}

The conservation of angular momentum represents the non-holonomic constraint in this system and couples $\dot{\phi}$ to its internal degrees of freedom $\tilde{a}$ and $\theta$, through the relation:
\begin{equation*}
    \dot{\phi} =  - \frac{b^2}{b^2+(L+a)^2} \; \dot{\theta}  = \tilde{a} \; \dot{\theta} \; .
\end{equation*}
This equation can be used to re-express the Lagrangian $\mathcal{L}$ solely in terms of $(\tilde{a}, \theta, \dot{\tilde{a}}, \dot{\theta})$. Note that the variable $a$ represents the displacements of the spring $k_1$ away from its rest length. This has motivate us to constrain the allowed values of $a$ to $a \in \; ]-L,L[$. Consequently, $\tilde{a}$ is bounded to the range $\tilde{a} \in \; ]-1,-1/(1+4L^2/b^2)[$. Hence, for the choice of parameters $L=b$, $\tilde{a}$ can only take values over the finite domain $\tilde{a} \in \;  ]-1,-1/5[$.

Proceeding in terms of the new variable $\tilde{a}$, the reduced Lagrangian reads:
\begin{equation}
    \begin{split}
        \mathcal{L}_{\text{red}} &= \frac{m b^2}{2} \bigg( \frac{\dot{\tilde{a}}^2}{4|\tilde{a}|^3(1-|\tilde{a}|)} + \dot{\theta}^2 (1-|\tilde{a}|) \bigg) + V_{12}(\tilde{a}) \\ 
        &+  V_{23}(\tilde{a},\theta) \; ,
    \end{split}
    \label{Lagrangian_shapespace}
\end{equation}

with
\begin{equation}
    \begin{split}
    & V_{12}(\tilde{a}) = \frac{k_1b^2}{2} \Bigg( \sqrt{\frac{1-|\tilde{a}|}{|\tilde{a}|}} - \frac{L}{b^2} \Bigg)^2 \; , \\
    & V_{23}(\tilde{a},\theta) = \frac{k_2b^2}{2} \Bigg( \sqrt{\frac{1}{|\tilde{a}|} -2 \cos(\theta) \sqrt{\frac{1-|\tilde{a}|}{|\tilde{a}|}} } - \frac{L}{b}\Bigg)^2 \; .
    \end{split}
\end{equation}
In eq.(\ref{Lagrangian_shapespace}), we have used the fact that $\tilde{a}$ is always negative to replace $-\tilde{a}$ by its modulus, $|\tilde{a}|$. The Hamiltonian can then be defined through the usual Legendre transformation~\cite{goldstein}:
\begin{equation}
    \begin{split}
        &\mathcal{H}_{\text{red}} (\mathbf{q},\mathbf{p}) = \sum^2_{i=1} p_i \dot{q}_i - \mathcal{L}_{\text{red}} (\mathbf{q},\dot{\mathbf{q}}) \\
        & p_i = \frac{\partial \mathcal{L}_{\text{red} }}{\partial q_i} \; , \; \mathbf{q} = (q_1,q_2)= (\tilde{a},\theta) \;,\\ 
        & \mathbf{p} = (p_1,p_2)= (p_{\tilde{a}},p_{\theta})   \; ,
    \end{split}
    \label{Legendre}
\end{equation}

which yields:
\begin{equation}
   \begin{split}
        \mathcal{H}_{red} &= \frac{1}{2mb^2} \Bigg( 4 \tilde{a}^2 (1-|\tilde{a}|) p^2_{\tilde{a}} + \frac{p^2_\theta}{1-|\tilde{a}|} \Bigg) + V_{12}(\tilde{a}) \\ 
        &+ V_{23} (\tilde{a},\theta) \; .
    \end{split}
\end{equation}

\section*{References}
\bibliography{RefChaticHam}

\providecommand{\noopsort}[1]{}\providecommand{\singleletter}[1]{#1}%
\begin{thebibliography}{43}%
\makeatletter
\providecommand \@ifxundefined [1]{%
 \@ifx{#1\undefined}
}%
\providecommand \@ifnum [1]{%
 \ifnum #1\expandafter \@firstoftwo
 \else \expandafter \@secondoftwo
 \fi
}%
\providecommand \@ifx [1]{%
 \ifx #1\expandafter \@firstoftwo
 \else \expandafter \@secondoftwo
 \fi
}%
\providecommand \natexlab [1]{#1}%
\providecommand \enquote  [1]{``#1''}%
\providecommand \bibnamefont  [1]{#1}%
\providecommand \bibfnamefont [1]{#1}%
\providecommand \citenamefont [1]{#1}%
\providecommand \href@noop [0]{\@secondoftwo}%
\providecommand \href [0]{\begingroup \@sanitize@url \@href}%
\providecommand \@href[1]{\@@startlink{#1}\@@href}%
\providecommand \@@href[1]{\endgroup#1\@@endlink}%
\providecommand \@sanitize@url [0]{\catcode `\\12\catcode `\$12\catcode `\&12\catcode `\#12\catcode `\^12\catcode `\_12\catcode `\%12\relax}%
\providecommand \@@startlink[1]{}%
\providecommand \@@endlink[0]{}%
\providecommand \url  [0]{\begingroup\@sanitize@url \@url }%
\providecommand \@url [1]{\endgroup\@href {#1}{\urlprefix }}%
\providecommand \urlprefix  [0]{URL }%
\providecommand \Eprint [0]{\href }%
\providecommand \doibase [0]{http://dx.doi.org/}%
\providecommand \selectlanguage [0]{\@gobble}%
\providecommand \bibinfo  [0]{\@secondoftwo}%
\providecommand \bibfield  [0]{\@secondoftwo}%
\providecommand \translation [1]{[#1]}%
\providecommand \BibitemOpen [0]{}%
\providecommand \bibitemStop [0]{}%
\providecommand \bibitemNoStop [0]{.\EOS\space}%
\providecommand \EOS [0]{\spacefactor3000\relax}%
\providecommand \BibitemShut  [1]{\csname bibitem#1\endcsname}%
\let\auto@bib@innerbib\@empty
\bibitem [{\citenamefont {McCauley}(1997)}]{mc_cauley}%
  \BibitemOpen
  \bibfield  {author} {\bibinfo {author} {\bibfnamefont {J.~L.}\ \bibnamefont {McCauley}},\ }\href@noop {} {\emph {\bibinfo {title} {Classical Mechanics: Transformations, Flows, Integrable and Chaotic Dynamics}}}\ (\bibinfo  {publisher} {Cambridge University Press},\ \bibinfo {year} {1997})\BibitemShut {NoStop}%
\bibitem [{\citenamefont {Goldstein}, \citenamefont {Poole},\ and\ \citenamefont {Safko}(2002)}]{goldstein}%
  \BibitemOpen
  \bibfield  {author} {\bibinfo {author} {\bibfnamefont {H.}~\bibnamefont {Goldstein}}, \bibinfo {author} {\bibfnamefont {C.}~\bibnamefont {Poole}}, \ and\ \bibinfo {author} {\bibfnamefont {J.}~\bibnamefont {Safko}},\ }\href@noop {} {\emph {\bibinfo {title} {Classical mechanics}}}\ (\bibinfo  {publisher} {Addison Wesley},\ \bibinfo {year} {2002})\BibitemShut {NoStop}%
\bibitem [{\citenamefont {Shapere}\ and\ \citenamefont {Wilczek}(1989)}]{shap_wilc}%
  \BibitemOpen
  \bibfield  {author} {\bibinfo {author} {\bibfnamefont {A.}~\bibnamefont {Shapere}}\ and\ \bibinfo {author} {\bibfnamefont {F.}~\bibnamefont {Wilczek}},\ }\bibfield  {title} {\enquote {\bibinfo {title} {Geometry of self-propulsion at low {R}eynolds number},}\ }\href@noop {} {\bibfield  {journal} {\bibinfo  {journal} {J. Fluid Mech.}\ }\textbf {\bibinfo {volume} {198}},\ \bibinfo {pages} {557--585} (\bibinfo {year} {1989})}\BibitemShut {NoStop}%
\bibitem [{\citenamefont {Berry}(1984)}]{berry_phase1}%
  \BibitemOpen
  \bibfield  {author} {\bibinfo {author} {\bibfnamefont {M.~V.}\ \bibnamefont {Berry}},\ }\bibfield  {title} {\enquote {\bibinfo {title} {Quantal phase factors accompanying adiabatic changes},}\ }\href@noop {} {\bibfield  {journal} {\bibinfo  {journal} {Proc. R. Soc. Lond. A}\ }\textbf {\bibinfo {volume} {392}},\ \bibinfo {pages} {45--57} (\bibinfo {year} {1984})}\BibitemShut {NoStop}%
\bibitem [{\citenamefont {Wilczek}\ and\ \citenamefont {Shapere}(1989)}]{wsgeometric}%
  \BibitemOpen
  \bibfield  {author} {\bibinfo {author} {\bibfnamefont {F.}~\bibnamefont {Wilczek}}\ and\ \bibinfo {author} {\bibfnamefont {A.}~\bibnamefont {Shapere}},\ }\href@noop {} {\emph {\bibinfo {title} {Geometric phases in physics}}},\ Vol.~\bibinfo {volume} {5}\ (\bibinfo  {publisher} {World Scientific},\ \bibinfo {year} {1989})\BibitemShut {NoStop}%
\bibitem [{\citenamefont {Wilczek}\ and\ \citenamefont {Zee}(1984)}]{WilcZee}%
  \BibitemOpen
  \bibfield  {author} {\bibinfo {author} {\bibfnamefont {F.}~\bibnamefont {Wilczek}}\ and\ \bibinfo {author} {\bibfnamefont {A.}~\bibnamefont {Zee}},\ }\bibfield  {title} {\enquote {\bibinfo {title} {Appearance of gauge structure in simple dynamical systems},}\ }\href@noop {} {\bibfield  {journal} {\bibinfo  {journal} {Phys. Rev. Lett.}\ }\textbf {\bibinfo {volume} {52}},\ \bibinfo {pages} {2111} (\bibinfo {year} {1984})}\BibitemShut {NoStop}%
\bibitem [{\citenamefont {Montgomery}(1993)}]{montgomery}%
  \BibitemOpen
  \bibfield  {author} {\bibinfo {author} {\bibfnamefont {R.}~\bibnamefont {Montgomery}},\ }\bibfield  {title} {\enquote {\bibinfo {title} {Gauge theory of the falling cat},}\ }\href@noop {} {\bibfield  {journal} {\bibinfo  {journal} {Fields Inst. Comm.}\ }\textbf {\bibinfo {volume} {1}},\ \bibinfo {pages} {193--218} (\bibinfo {year} {1993})}\BibitemShut {NoStop}%
\bibitem [{\citenamefont {Littlejohn}\ and\ \citenamefont {Reinsch}(1997)}]{molgeophase}%
  \BibitemOpen
  \bibfield  {author} {\bibinfo {author} {\bibfnamefont {R.~G.}\ \bibnamefont {Littlejohn}}\ and\ \bibinfo {author} {\bibfnamefont {M.}~\bibnamefont {Reinsch}},\ }\bibfield  {title} {\enquote {\bibinfo {title} {Gauge fields in the separation of rotations andinternal motions in the n-body problem},}\ }\href@noop {} {\bibfield  {journal} {\bibinfo  {journal} {Rev. Mod. Phys.}\ }\textbf {\bibinfo {volume} {69}},\ \bibinfo {pages} {213--276} (\bibinfo {year} {1997})}\BibitemShut {NoStop}%
\bibitem [{\citenamefont {Zwanziger}, \citenamefont {Koenig},\ and\ \citenamefont {Pines}(1990)}]{berry_phase2}%
  \BibitemOpen
  \bibfield  {author} {\bibinfo {author} {\bibfnamefont {J.~W.}\ \bibnamefont {Zwanziger}}, \bibinfo {author} {\bibfnamefont {M.}~\bibnamefont {Koenig}}, \ and\ \bibinfo {author} {\bibfnamefont {A.}~\bibnamefont {Pines}},\ }\bibfield  {title} {\enquote {\bibinfo {title} {Berry's phase},}\ }\href@noop {} {\bibfield  {journal} {\bibinfo  {journal} {Annu. Rev. Phys. Chem.}\ }\textbf {\bibinfo {volume} {41}},\ \bibinfo {pages} {601--646} (\bibinfo {year} {1990})}\BibitemShut {NoStop}%
\bibitem [{\citenamefont {Berry}(1987)}]{light_pol}%
  \BibitemOpen
  \bibfield  {author} {\bibinfo {author} {\bibfnamefont {M.~V.}\ \bibnamefont {Berry}},\ }\bibfield  {title} {\enquote {\bibinfo {title} {The adiabatic phase and {P}ancharatnam's phase for polarized light},}\ }\href@noop {} {\bibfield  {journal} {\bibinfo  {journal} {J. Mod. Opt.}\ }\textbf {\bibinfo {volume} {34}},\ \bibinfo {pages} {1401--1407} (\bibinfo {year} {1987})}\BibitemShut {NoStop}%
\bibitem [{\citenamefont {Putterman}\ and\ \citenamefont {Raz}(2008)}]{square_cat}%
  \BibitemOpen
  \bibfield  {author} {\bibinfo {author} {\bibfnamefont {E.}~\bibnamefont {Putterman}}\ and\ \bibinfo {author} {\bibfnamefont {O.}~\bibnamefont {Raz}},\ }\bibfield  {title} {\enquote {\bibinfo {title} {The square cat},}\ }\href@noop {} {\bibfield  {journal} {\bibinfo  {journal} {Am. J. Phys.}\ }\textbf {\bibinfo {volume} {76}},\ \bibinfo {pages} {1040--1044} (\bibinfo {year} {2008})}\BibitemShut {NoStop}%
\bibitem [{\citenamefont {Montgomery}(2015)}]{montgomery2}%
  \BibitemOpen
  \bibfield  {author} {\bibinfo {author} {\bibfnamefont {R.}~\bibnamefont {Montgomery}},\ }\bibfield  {title} {\enquote {\bibinfo {title} {The three-body problem and the shape sphere},}\ }\href@noop {} {\bibfield  {journal} {\bibinfo  {journal} {Am. Math. Mon.}\ }\textbf {\bibinfo {volume} {122}},\ \bibinfo {pages} {299--321} (\bibinfo {year} {2015})}\BibitemShut {NoStop}%
\bibitem [{\citenamefont {S.~Katz}\ and\ \citenamefont {Efrati}(2019)}]{ori1}%
  \BibitemOpen
  \bibfield  {author} {\bibinfo {author} {\bibfnamefont {O.}~\bibnamefont {S.~Katz}}\ and\ \bibinfo {author} {\bibfnamefont {E.}~\bibnamefont {Efrati}},\ }\bibfield  {title} {\enquote {\bibinfo {title} {Self-driven fractional rotational diffusion of the harmonic three-mass system},}\ }\href@noop {} {\bibfield  {journal} {\bibinfo  {journal} {Phys. Rev. Lett.}\ }\textbf {\bibinfo {volume} {122}},\ \bibinfo {pages} {024102} (\bibinfo {year} {2019})}\BibitemShut {NoStop}%
\bibitem [{\citenamefont {S.~Katz}\ and\ \citenamefont {Efrati}(2020)}]{ori2}%
  \BibitemOpen
  \bibfield  {author} {\bibinfo {author} {\bibfnamefont {O.}~\bibnamefont {S.~Katz}}\ and\ \bibinfo {author} {\bibfnamefont {E.}~\bibnamefont {Efrati}},\ }\bibfield  {title} {\enquote {\bibinfo {title} {Regular regimes of the harmonic three-mass system},}\ }\href@noop {} {\bibfield  {journal} {\bibinfo  {journal} {Phys. Rev. E}\ }\textbf {\bibinfo {volume} {101}},\ \bibinfo {pages} {032211} (\bibinfo {year} {2020})}\BibitemShut {NoStop}%
\bibitem [{\citenamefont {Iwai}(1987)}]{iwai}%
  \BibitemOpen
  \bibfield  {author} {\bibinfo {author} {\bibfnamefont {T.}~\bibnamefont {Iwai}},\ }\bibfield  {title} {\enquote {\bibinfo {title} {A gauge theory for the quantum planar three‐body problem},}\ }\href@noop {} {\bibfield  {journal} {\bibinfo  {journal} {J. Math. Phys.}\ }\textbf {\bibinfo {volume} {28}},\ \bibinfo {pages} {964--974} (\bibinfo {year} {1987})}\BibitemShut {NoStop}%
\bibitem [{\citenamefont {L{\'e}vy}(1948)}]{levy}%
  \BibitemOpen
  \bibfield  {author} {\bibinfo {author} {\bibfnamefont {P.}~\bibnamefont {L{\'e}vy}},\ }\href@noop {} {\emph {\bibinfo {title} {Processus Stochastiques et Mouvements Browniens}}}\ (\bibinfo  {publisher} {Paris},\ \bibinfo {year} {1948})\BibitemShut {NoStop}%
\bibitem [{\citenamefont {Edwards}(1967)}]{edwards_stat}%
  \BibitemOpen
  \bibfield  {author} {\bibinfo {author} {\bibfnamefont {S.~F.}\ \bibnamefont {Edwards}},\ }\bibfield  {title} {\enquote {\bibinfo {title} {Statistical mechanics with topological constraints: I},}\ }\href@noop {} {\bibfield  {journal} {\bibinfo  {journal} {Proc. Phys. Soc.}\ }\textbf {\bibinfo {volume} {91}},\ \bibinfo {pages} {513--519} (\bibinfo {year} {1967})}\BibitemShut {NoStop}%
\bibitem [{\citenamefont {Brereton}\ and\ \citenamefont {Butler}(1987)}]{brereton1987topological}%
  \BibitemOpen
  \bibfield  {author} {\bibinfo {author} {\bibfnamefont {M.~G.}\ \bibnamefont {Brereton}}\ and\ \bibinfo {author} {\bibfnamefont {C.}~\bibnamefont {Butler}},\ }\bibfield  {title} {\enquote {\bibinfo {title} {A topological problem in polymer physics: configurational and mechanical properties of a random walk enclosing a constant area},}\ }\href@noop {} {\bibfield  {journal} {\bibinfo  {journal} {J. Phys. A : Math. Gen.}\ }\textbf {\bibinfo {volume} {20}},\ \bibinfo {pages} {3955--3968} (\bibinfo {year} {1987})}\BibitemShut {NoStop}%
\bibitem [{\citenamefont {Khandekar}\ and\ \citenamefont {Wiegel}(1988)}]{khandekar1988distribution}%
  \BibitemOpen
  \bibfield  {author} {\bibinfo {author} {\bibfnamefont {D.~C.}\ \bibnamefont {Khandekar}}\ and\ \bibinfo {author} {\bibfnamefont {F.~W.}\ \bibnamefont {Wiegel}},\ }\bibfield  {title} {\enquote {\bibinfo {title} {Distribution of the area enclosed by a plane random walk},}\ }\href@noop {} {\bibfield  {journal} {\bibinfo  {journal} {J. Phys. A: Math. Gen.}\ }\textbf {\bibinfo {volume} {21}},\ \bibinfo {pages} {L563--L566} (\bibinfo {year} {1988})}\BibitemShut {NoStop}%
\bibitem [{\citenamefont {Duplantier}(1989)}]{duplantier1989areas}%
  \BibitemOpen
  \bibfield  {author} {\bibinfo {author} {\bibfnamefont {B.}~\bibnamefont {Duplantier}},\ }\bibfield  {title} {\enquote {\bibinfo {title} {Areas of planar {B}rownian curves},}\ }\href@noop {} {\bibfield  {journal} {\bibinfo  {journal} {J. Phys. A: Math. Gen.}\ }\textbf {\bibinfo {volume} {22}},\ \bibinfo {pages} {3033--3048} (\bibinfo {year} {1989})}\BibitemShut {NoStop}%
\bibitem [{\citenamefont {Desbois}\ and\ \citenamefont {Comtet}(1992)}]{desbois1992algebraic}%
  \BibitemOpen
  \bibfield  {author} {\bibinfo {author} {\bibfnamefont {J.}~\bibnamefont {Desbois}}\ and\ \bibinfo {author} {\bibfnamefont {A.}~\bibnamefont {Comtet}},\ }\bibfield  {title} {\enquote {\bibinfo {title} {Algebraic areas enclosed by {B}rownian curves on bounded domains},}\ }\href@noop {} {\bibfield  {journal} {\bibinfo  {journal} {J. Phys. A: Math. Gen.}\ }\textbf {\bibinfo {volume} {25}},\ \bibinfo {pages} {3097--3102} (\bibinfo {year} {1992})}\BibitemShut {NoStop}%
\bibitem [{\citenamefont {Marsden}, \citenamefont {Ratiu},\ and\ \citenamefont {Scheurle}(2000)}]{symplectic-reduction}%
  \BibitemOpen
  \bibfield  {author} {\bibinfo {author} {\bibfnamefont {J.~E.}\ \bibnamefont {Marsden}}, \bibinfo {author} {\bibfnamefont {T.~S.}\ \bibnamefont {Ratiu}}, \ and\ \bibinfo {author} {\bibfnamefont {J.}~\bibnamefont {Scheurle}},\ }\bibfield  {title} {\enquote {\bibinfo {title} {Reduction theory and the {L}agrange - {R}outh equations},}\ }\href@noop {} {\bibfield  {journal} {\bibinfo  {journal} {J. Math. Phys.}\ }\textbf {\bibinfo {volume} {41}},\ \bibinfo {pages} {3379--3429} (\bibinfo {year} {2000})}\BibitemShut {NoStop}%
\bibitem [{\citenamefont {Ott}(2002)}]{ott}%
  \BibitemOpen
  \bibfield  {author} {\bibinfo {author} {\bibfnamefont {E.}~\bibnamefont {Ott}},\ }\href@noop {} {\emph {\bibinfo {title} {Chaos in dynamical systems}}}\ (\bibinfo  {publisher} {Cambridge University Press, Cambridge, UK},\ \bibinfo {year} {2002})\BibitemShut {NoStop}%
\bibitem [{\citenamefont {Strogatz}(2015)}]{strogratz}%
  \BibitemOpen
  \bibfield  {author} {\bibinfo {author} {\bibfnamefont {S.}~\bibnamefont {Strogatz}},\ }\href@noop {} {\emph {\bibinfo {title} {Nonlinear {D}ynamics and {C}haos: {W}ith {A}pplications to {P}hysics, {B}iology, {C}hemistry, and {E}ngineering}}}\ (\bibinfo  {publisher} {CRC Press},\ \bibinfo {year} {2015})\BibitemShut {NoStop}%
\bibitem [{\citenamefont {Silva}, \citenamefont {Ben~Av},\ and\ \citenamefont {Efrati}(2021)}]{symplectic_int}%
  \BibitemOpen
  \bibfield  {author} {\bibinfo {author} {\bibfnamefont {A.}~\bibnamefont {Silva}}, \bibinfo {author} {\bibfnamefont {E.}~\bibnamefont {Ben~Av}}, \ and\ \bibinfo {author} {\bibfnamefont {E.}~\bibnamefont {Efrati}},\ }\bibfield  {title} {\enquote {\bibinfo {title} {Explicit, time-reversible and symplectic integrator for hamiltonians in isotropic uniformly curved geometries},}\ }\href@noop {} {\bibfield  {journal} {\bibinfo  {journal} {arXiv:2104.10908v1 [math.NA]}\ } (\bibinfo {year} {2021})}\BibitemShut {NoStop}%
\bibitem [{\citenamefont {Maggs}(2001)}]{maggs1}%
  \BibitemOpen
  \bibfield  {author} {\bibinfo {author} {\bibfnamefont {A.~C.}\ \bibnamefont {Maggs}},\ }\bibfield  {title} {\enquote {\bibinfo {title} {Writhing geometry at finite temperature: {R}andom walks and geometric phases for stiff polymers},}\ }\href@noop {} {\bibfield  {journal} {\bibinfo  {journal} {J. Chem. Phys.}\ }\textbf {\bibinfo {volume} {114}},\ \bibinfo {pages} {5888--5896} (\bibinfo {year} {2001})}\BibitemShut {NoStop}%
\bibitem [{\citenamefont {Maggs}\ and\ \citenamefont {Rossetto}(2001)}]{maggs2}%
  \BibitemOpen
  \bibfield  {author} {\bibinfo {author} {\bibfnamefont {A.~C.}\ \bibnamefont {Maggs}}\ and\ \bibinfo {author} {\bibfnamefont {V.}~\bibnamefont {Rossetto}},\ }\bibfield  {title} {\enquote {\bibinfo {title} {Writhing photons and {B}erry phases in polarized multiple scattering},}\ }\href@noop {} {\bibfield  {journal} {\bibinfo  {journal} {Phys. Rev. Lett.}\ }\textbf {\bibinfo {volume} {87}},\ \bibinfo {pages} {253901} (\bibinfo {year} {2001})}\BibitemShut {NoStop}%
\bibitem [{\citenamefont {Sinha}\ and\ \citenamefont {Samuel}(1994)}]{sinha1994brownian}%
  \BibitemOpen
  \bibfield  {author} {\bibinfo {author} {\bibfnamefont {S.}~\bibnamefont {Sinha}}\ and\ \bibinfo {author} {\bibfnamefont {J.}~\bibnamefont {Samuel}},\ }\bibfield  {title} {\enquote {\bibinfo {title} {Brownian motion and magnetism},}\ }\href@noop {} {\bibfield  {journal} {\bibinfo  {journal} {Phys. Rev. B}\ }\textbf {\bibinfo {volume} {50}},\ \bibinfo {pages} {13871} (\bibinfo {year} {1994})}\BibitemShut {NoStop}%
\bibitem [{\citenamefont {Antoine}\ \emph {et~al.}(1991)\citenamefont {Antoine}, \citenamefont {Comtet}, \citenamefont {Desbois},\ and\ \citenamefont {Ouvry}}]{brown_sphere}%
  \BibitemOpen
  \bibfield  {author} {\bibinfo {author} {\bibfnamefont {M.}~\bibnamefont {Antoine}}, \bibinfo {author} {\bibfnamefont {A.}~\bibnamefont {Comtet}}, \bibinfo {author} {\bibfnamefont {J.}~\bibnamefont {Desbois}}, \ and\ \bibinfo {author} {\bibfnamefont {S.}~\bibnamefont {Ouvry}},\ }\bibfield  {title} {\enquote {\bibinfo {title} {Magnetic fields and {B}rownian motion on the 2-sphere},}\ }\href@noop {} {\bibfield  {journal} {\bibinfo  {journal} {J. Phys. A: Math. Gen.}\ }\textbf {\bibinfo {volume} {24}},\ \bibinfo {pages} {2581--2586} (\bibinfo {year} {1991})}\BibitemShut {NoStop}%
\bibitem [{\citenamefont {Wiegel}(1986)}]{wiegel1986}%
  \BibitemOpen
  \bibfield  {author} {\bibinfo {author} {\bibfnamefont {F.~W.}\ \bibnamefont {Wiegel}},\ }\href@noop {} {\emph {\bibinfo {title} {Introduction to path-integral methods in physics and polymer science}}}\ (\bibinfo  {publisher} {World Scientific, Singapore},\ \bibinfo {year} {1986})\BibitemShut {NoStop}%
\bibitem [{\citenamefont {Karr}(1993)}]{probability}%
  \BibitemOpen
  \bibfield  {author} {\bibinfo {author} {\bibfnamefont {A.~F.}\ \bibnamefont {Karr}},\ }\href@noop {} {\emph {\bibinfo {title} {Probability}}},\ Springer Texts in Statistics.\ (\bibinfo  {publisher} {Springer-Verlag, New York, NY.},\ \bibinfo {year} {1993})\BibitemShut {NoStop}%
\bibitem [{\citenamefont {Khandekar}\ and\ \citenamefont {Lawande}(1986)}]{khandekar}%
  \BibitemOpen
  \bibfield  {author} {\bibinfo {author} {\bibfnamefont {D.~C.}\ \bibnamefont {Khandekar}}\ and\ \bibinfo {author} {\bibfnamefont {S.~V.}\ \bibnamefont {Lawande}},\ }\bibfield  {title} {\enquote {\bibinfo {title} {Feynman path integrals: {s}ome exact results and applications},}\ }\href@noop {} {\bibfield  {journal} {\bibinfo  {journal} {Phys. Rep.}\ }\textbf {\bibinfo {volume} {137}},\ \bibinfo {pages} {115--229} (\bibinfo {year} {1986})}\BibitemShut {NoStop}%
\bibitem [{\citenamefont {Feynman}\ and\ \citenamefont {Hibbs}(1965)}]{feynman}%
  \BibitemOpen
  \bibfield  {author} {\bibinfo {author} {\bibfnamefont {R.~P.}\ \bibnamefont {Feynman}}\ and\ \bibinfo {author} {\bibfnamefont {A.~R.}\ \bibnamefont {Hibbs}},\ }\href@noop {} {\emph {\bibinfo {title} {Quantum mechanics and path integrals}}}\ (\bibinfo  {publisher} {New York: McGraw-Hill},\ \bibinfo {year} {1965})\BibitemShut {NoStop}%
\bibitem [{\citenamefont {de~Wijn}\ and\ \citenamefont {Fasolino}(2009)}]{wijn}%
  \BibitemOpen
  \bibfield  {author} {\bibinfo {author} {\bibfnamefont {A.~S.}\ \bibnamefont {de~Wijn}}\ and\ \bibinfo {author} {\bibfnamefont {A.}~\bibnamefont {Fasolino}},\ }\bibfield  {title} {\enquote {\bibinfo {title} {Relating chaos to deterministic diffusion of a molecule adsorbed on a surface},}\ }\href@noop {} {\bibfield  {journal} {\bibinfo  {journal} {J. Phys.: Condens. Matter}\ }\textbf {\bibinfo {volume} {21}},\ \bibinfo {pages} {264002} (\bibinfo {year} {2009})}\BibitemShut {NoStop}%
\bibitem [{Note1()}]{Note1}%
  \BibitemOpen
  \bibinfo {note} {The straightforward example reads $\protect \tilde {x}^2=x^2,\protect \tilde {x}^1=\DOTSI \intop \ilimits@ f(x^1,x^2)dx^1 $.}\BibitemShut {Stop}%
\bibitem [{\citenamefont {Moser}(1965)}]{moser}%
  \BibitemOpen
  \bibfield  {author} {\bibinfo {author} {\bibfnamefont {J.}~\bibnamefont {Moser}},\ }\bibfield  {title} {\enquote {\bibinfo {title} {On the volume elements on a manifold},}\ }\href@noop {} {\bibfield  {journal} {\bibinfo  {journal} {Trans. Am. Math. Soc.}\ }\textbf {\bibinfo {volume} {120}},\ \bibinfo {pages} {286--294} (\bibinfo {year} {1965})}\BibitemShut {NoStop}%
\bibitem [{\citenamefont {Skokos}(2010)}]{skokos2010lyapunov}%
  \BibitemOpen
  \bibfield  {author} {\bibinfo {author} {\bibfnamefont {C.}~\bibnamefont {Skokos}},\ }\enquote {\bibinfo {title} {The {L}yapunov characteristic exponents and their computation},}\ in\ \href@noop {} {\emph {\bibinfo {booktitle} {Dynamics of Small Solar System Bodies and Exoplanets}}},\ \bibinfo {editor} {edited by\ \bibinfo {editor} {\bibfnamefont {J.~J.}\ \bibnamefont {Souchay}}}\ (\bibinfo  {publisher} {Springer Berlin Heidelberg},\ \bibinfo {address} {Berlin, Heidelberg},\ \bibinfo {year} {2010})\ pp.\ \bibinfo {pages} {63--135}\BibitemShut {NoStop}%
\bibitem [{\citenamefont {Lowenstein}(2012)}]{lowenstein2012essentials}%
  \BibitemOpen
  \bibfield  {author} {\bibinfo {author} {\bibfnamefont {J.~H.}\ \bibnamefont {Lowenstein}},\ }\href@noop {} {\emph {\bibinfo {title} {Essentials of Hamiltonian dynamics}}}\ (\bibinfo  {publisher} {Cambridge University Press, Cambridge, UK},\ \bibinfo {year} {2012})\BibitemShut {NoStop}%
\bibitem [{\citenamefont {Pikovsky}\ and\ \citenamefont {Politi}(2016)}]{book_lyapunov_1}%
  \BibitemOpen
  \bibfield  {author} {\bibinfo {author} {\bibfnamefont {A.}~\bibnamefont {Pikovsky}}\ and\ \bibinfo {author} {\bibfnamefont {A.}~\bibnamefont {Politi}},\ }\href@noop {} {\emph {\bibinfo {title} {Lyapunov exponents: a tool to explore complex dynamics}}}\ (\bibinfo  {publisher} {Cambridge University Press},\ \bibinfo {year} {2016})\BibitemShut {NoStop}%
\bibitem [{\citenamefont {Contopoulos}, \citenamefont {Galgani},\ and\ \citenamefont {Giorgilli}(1978)}]{contopoulos1978number}%
  \BibitemOpen
  \bibfield  {author} {\bibinfo {author} {\bibfnamefont {G.}~\bibnamefont {Contopoulos}}, \bibinfo {author} {\bibfnamefont {L.}~\bibnamefont {Galgani}}, \ and\ \bibinfo {author} {\bibfnamefont {A.}~\bibnamefont {Giorgilli}},\ }\bibfield  {title} {\enquote {\bibinfo {title} {On the number of isolating integrals in hamiltonian systems},}\ }\href@noop {} {\bibfield  {journal} {\bibinfo  {journal} {Phys. Rev. A}\ }\textbf {\bibinfo {volume} {18}},\ \bibinfo {pages} {1183} (\bibinfo {year} {1978})}\BibitemShut {NoStop}%
\bibitem [{\citenamefont {Ramasubramanian}\ and\ \citenamefont {Sriram}(2000)}]{numericalLyapunov}%
  \BibitemOpen
  \bibfield  {author} {\bibinfo {author} {\bibfnamefont {K.}~\bibnamefont {Ramasubramanian}}\ and\ \bibinfo {author} {\bibfnamefont {M.}~\bibnamefont {Sriram}},\ }\bibfield  {title} {\enquote {\bibinfo {title} {A comparative study of computation of {L}yapunov spectra with different algorithms},}\ }\href@noop {} {\bibfield  {journal} {\bibinfo  {journal} {Phys. D: Nonlinear Phenom.}\ }\textbf {\bibinfo {volume} {139}},\ \bibinfo {pages} {72--86} (\bibinfo {year} {2000})}\BibitemShut {NoStop}%
\bibitem [{\citenamefont {Desbois}(2015)}]{desbois2015}%
  \BibitemOpen
  \bibfield  {author} {\bibinfo {author} {\bibfnamefont {J.}~\bibnamefont {Desbois}},\ }\bibfield  {title} {\enquote {\bibinfo {title} {Algebraic area enclosed by random walks on a lattice},}\ }\href@noop {} {\bibfield  {journal} {\bibinfo  {journal} {J. Phys. A: Math. Theor.}\ }\textbf {\bibinfo {volume} {48}},\ \bibinfo {pages} {425001} (\bibinfo {year} {2015})}\BibitemShut {NoStop}%
\bibitem [{\citenamefont {Mashkevich}\ and\ \citenamefont {Ouvry}(2009)}]{mashkevich2009area}%
  \BibitemOpen
  \bibfield  {author} {\bibinfo {author} {\bibfnamefont {S.}~\bibnamefont {Mashkevich}}\ and\ \bibinfo {author} {\bibfnamefont {S.}~\bibnamefont {Ouvry}},\ }\bibfield  {title} {\enquote {\bibinfo {title} {Area distribution of two-dimensional random walks on a square lattice},}\ }\href@noop {} {\bibfield  {journal} {\bibinfo  {journal} {J. Stat. Phys.}\ }\textbf {\bibinfo {volume} {137}},\ \bibinfo {pages} {71--78} (\bibinfo {year} {2009})}\BibitemShut {NoStop}%
\end{thebibliography}%

\end{document}